%% file: main.tex
\documentclass[sigconf,nonacm]{acmart}
\AtBeginDocument{%
  \providecommand\BibTeX{{%
    Bib\TeX}}}

\usepackage{amsmath}
\usepackage{mathtools}
\usepackage{amsfonts}
\usepackage{caption}
\usepackage{algorithmic}
\usepackage[ruled,vlined]{algorithm2e}
\usepackage{graphicx}
\usepackage{textcomp}
\usepackage{xcolor}
\usepackage{verbatim}
\usepackage{subcaption}
\usepackage{adjustbox}
\usepackage{booktabs}
\usepackage{multirow}
\usepackage{threeparttable, tablefootnote}
\usepackage{varwidth,xcolor}

\usepackage{tabularray}
\usepackage{ragged2e}
\usepackage{array}
\usepackage{bm}
\usepackage{mathrsfs}
\usepackage{tabularx}
\usepackage{makecell}
\usepackage{diagbox}

\newcommand{\llangle}{\langle\mkern-4.8mu\langle}
\newcommand{\rrangle}{\rangle\mkern-4.8mu\rangle}

\newcommand{\ignore}[1]{}
\newcommand{\fixme}[1]{\textcolor{red}{#1}}

\newcommand{\micro}{\mu}
\begin{document}
\newcolumntype{C}[1]{>{\centering\arraybackslash}p{#1}}
\title[Flash: A Hybrid Private Inference Protocol for Deep CNNs with High Accuracy and Low Latency on CPU]
{Flash: A Hybrid Private Inference Protocol for Deep CNNs\\with High Accuracy and Low Latency on CPU}


\ignore{\author{Hyeri Roh$^1$, Jinsu Yeo$^1$, Yeongil Ko$^2$, Gu-Yeon Wei$^3$, David Brooks$^2$, and Woo-Seok Choi$^1$}
\affiliation{%
  \institution{$^1$Dept. of ECE, ISRC, Seoul National University, Seoul, South Korea}
  \streetaddress{1 Gwanak-ro, Gwanak-gu}
  \city{$^2$Harvard University, Cambridge, MA}
  \country{USA}}
  \email{{hrroh,wooseokchoi}@snu.ac.kr}
}

\author{Hyeri Roh}
\affiliation{%
  \institution{Dept. of ECE, ISRC}
  \streetaddress{1 Gwanak-ro, Gwanak-gu}
  \city{Seoul National University}
  \country{}
  }
  \email{hrroh@snu.ac.kr}

\author{Jinsu Yeo}
\affiliation{%
  \institution{Seoul National University}
  \streetaddress{1 Gwanak-ro, Gwanak-gu}
  \city{Samsung Research}
  \country{}
  }
  \email{yjs720@snu.ac.kr}

\author{Yeongil Ko}
\affiliation{%
  \institution{Harvard University}
  \streetaddress{1 Gwanak-ro, Gwanak-gu}
  \city{now at Google}
  \country{}
  }
  \email{yeongil_ko@g.harvard.edu}

\author{Gu-Yeon Wei}
\affiliation{%
  \institution{}
  \streetaddress{1 Gwanak-ro, Gwanak-gu}
  \city{Harvard University}
  \country{}
  }
  \email{guyeon@seas.harvard.edu}  

\author{David Brooks}
\affiliation{%
  \institution{}
  \streetaddress{1 Gwanak-ro, Gwanak-gu}
  \city{Harvard University}
  \country{}
  }
  \email{dbrooks@g.harvard.edu}  

\author{Woo-Seok Choi}
\affiliation{%
  \institution{Dept. of ECE, ISRC}
  \streetaddress{1 Gwanak-ro, Gwanak-gu}
  \city{Seoul National University}
  \country{}
  }
  \email{wooseokchoi@snu.ac.kr}

\input{Contents/0_abstract}
\maketitle
\input{Contents/1_intro}
\input{Contents/2_background}
\input{Contents/3_conv}
\input{Contents/4_training}
\input{Contents/5_mpc}
\input{Contents/6_flash}

\input{Contents/7_results}
\input{Contents/9_conclusion}

\begin{acks}
    This work was supported by the New Faculty Startup Fund from Seoul National University.
\end{acks}

\bibliographystyle{ACM-Reference-Format}
\bibliography{main}
\input{Contents/appendix.tex}

\end{document}

%% file: Contents/0_abstract.tex
\begin{abstract}
This paper presents Flash, an optimized private inference (PI) hybrid protocol utilizing both homomorphic encryption (HE) and secure two-party computation (2PC), which can reduce the end-to-end PI latency for deep CNN models less than 1 minute with CPU.
To this end, first, Flash proposes a low-latency convolution algorithm built upon a fast slot rotation operation and a novel data encoding scheme, which results in {4-94}$\,\times$ performance gain over the state-of-the-art.
Second, to minimize the communication cost introduced by the standard nonlinear activation function ReLU,
Flash replaces all ReLUs with the polynomial $x^2+x$ and trains deep CNN models with the new training strategy, which improves the inference accuracy for CIFAR-10/100 and TinyImageNet by 16\,\% on average (up to 40\,\% for ResNet-32) compared to prior art.
Lastly, Flash proposes an efficient 2PC-based $x^2+x$ evaluation protocol that does not require any offline communication and reduces the total communication cost to process the activation layer by {84-196}\,$\times$ over the state of the art.
As a result, the end-to-end PI latency of Flash implemented on CPU is 0.02 minute for CIFAR-100 and 0.57 minute for TinyImageNet classification, while the total data communication is 0.07\,GB for CIFAR-100 and 0.22\,GB for TinyImageNet.
Flash improves the state-of-the-art PI by {16-45}\,$\times$ in latency and {84-196}\,$\times$ in communication cost.
Moreover, even for ImageNet, Flash can deliver the latency less than 1 minute on CPU with the total communication less than 1\,GB.
\end{abstract}

%% file: Contents/1_intro.tex
\section{Introduction}
\label{sec:intro}

\begin{figure*}[t]
\centering
\includegraphics[width=1.87\columnwidth]{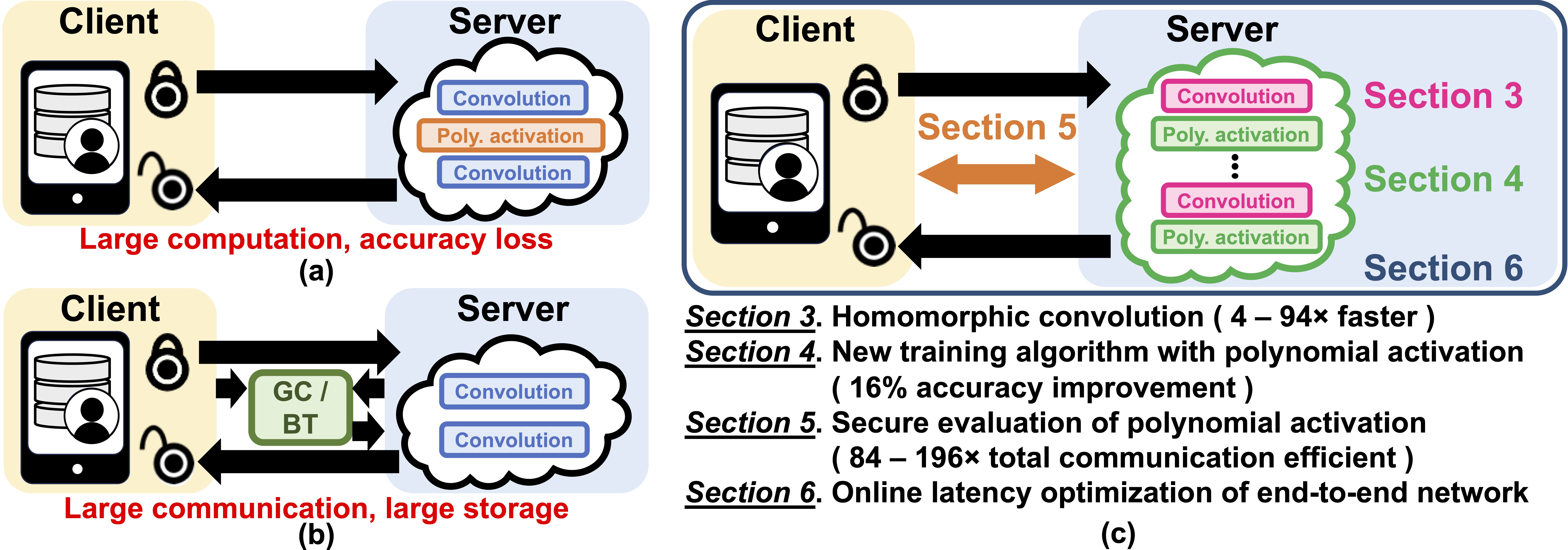} 
\vspace{-0.5em}
\caption{System design for PI protocol: (a) HE-based, (b) 2PC-based, and (c) overview of Flash with this paper's organization.}
\label{fig:overview}
\vspace{-0.5em}
\end{figure*}

With the recent advance in ML/AI applications, a growing number of services such as searching, recommendation, classification, translation, etc., have been and are being replaced with data-driven, cloud-based approaches that take advantage of service users' personal data in training and deploying stages.
While these services can provide huge benefits in many aspects of daily life,
the risk of undermining personal privacy is rapidly growing behind the scene~\cite{subashini2011survey,chen2012data}.
Although privacy issues prevail in both the training and inference phases in ML applications,
this paper focuses on privacy-preserving inference, or private inference (PI), 
where ML models in the cloud should provide inference without compromising the privacy of the client's data or the model itself.

To alleviate the privacy concerns, many techniques have been introduced in the literature~\cite{costan2016intel,satyanarayanan2017emergence,cormode2018privacy,choi2018guaranteeing,mohassel2017secureml,liu2017oblivious,gilad2016cryptonets,juvekar2018gazelle,mothukuri2021survey,li2020federated}.
However, each technique comes with different computing or communication costs and security guarantees.
Hardware-based solutions (trusted execution environments, or TEEs) like Intel's Software Guard Extensions (SGX)~\cite{costan2016intel} provide protected regions of memory called enclaves, 
which can be used to securely run ML inference without exposing data or the model to the rest of the system.
However, this technique is known to be vulnerable to side-channel attacks~\cite{nilsson2020survey}.
Differential privacy~\cite{dwork2014algorithmic,cormode2018privacy} provides a mathematical guarantee that the output of a function like an ML model does not reveal too much information about any single input,
but most of the time, it is used in the context of data analysis or model training~\cite{abadi2016deep}.
Homomorphic encryption (HE)~\cite{gentry2009fully} is a form of encryption that allows computation on ciphertexts, generating an encrypted result that, when decrypted, matches the result of the operations as if they had been performed on the plaintext.
In ML, HE enables a model to make predictions on encrypted data, and the result is then decrypted by the client without the server ever seeing the raw data.
Secure multiparty computation (MPC)~\cite{goldreich1998secure} allows multiple parties to collaboratively compute a function over their inputs while keeping those inputs private.
In ML, this implies that, with MPC, different entities can collaborate to train or infer from a model without exposing their individual data to each other.

Since privacy concerns continue to grow in significance,
there has been active research in building an efficient PI protocol~\cite{gilad2016cryptonets,hesamifard2017cryptodl,brutzkus2019low,lee2021high,juvekar2018gazelle,choi2022impala,rathee2020cryptflow2,huang2022cheetah} using either HE, 2PC, or both.
PI protocols based solely on HE (see Fig.~\ref{fig:overview}(a)) suffer from massive computational overhead, and the inference latency goes over 1 hour for ResNet-32 on CPU~\cite{lee2021high}.
Without the aid of hardware accelerators, HE alone would not be a viable option for PI.
In addition, ReLU in a model should be replaced with a polynomial activation function, which leads to model accuracy degradation~\cite{garimella2021sisyphus}.
On the other hand, 2PC-based PI (see Fig.~\ref{fig:overview}(b)) requires multiple rounds of communication and incurs huge communication costs, e.g., more than several GB for a single PI with ResNet-32~\cite{mishra2020delphi}.
Some parts of the protocol can be performed during the offline phase to minimize the online latency and communication cost,
but this requires large storage to keep the precomputed values, and the total communication cost per inference still remains the huge bottleneck.
Prior art such as \cite{juvekar2018gazelle,choi2022impala,huang2022cheetah} try to tackle this problem by combining HE and 2PC,
but their CPU implementations are either limited to shallow convolutional neural networks (CNNs) for CIFAR-10 classification or computation/communication-costly with deep CNNs for ImageNet inference.

In order to make PI more practical, it is required to further reduce the computation and communication overhead incurred by HE and 2PC by orders of magnitude.
We propose \textit{Flash}, a system-wise optimized hybrid HE/2PC PI protocol that can reduce the end-to-end inference latency for deep CNN models with high accuracy to less than 1 minute on CPU.
To this end, we solve three fundamental problems existing in prior art:
1) large computational overhead for convolution due to HE,
2) model accuracy degradation caused by replacing ReLU with a polynomial activation function,
and 3) large communication costs per inference to process nonlinear activation layer due to 2PC.
Specifically, this paper makes the following contributions (see Fig.~\ref{fig:overview}(c)):
\begin{itemize}
    \item A low-latency convolution algorithm along with a fast slot rotation operation and a novel data vector encoding scheme is proposed, which provides up to {94}\,$\times$ performance improvement over the state-of-the-art.
    \item We present a new training algorithm for deep CNNs with the polynomial activation function, $x^2 + x$, and show that the inference accuracy for CIFAR-10/100 and TinyImageNet improves by 16\,\% on average (up to 40\,\% in ResNet-32) compared to the state-of-the-art.
    \item A low-latency communication-efficient 2PC-based $x^2+x$ evaluation protocol is proposed, which does not require any offline communication. This leads to the total communication cost saving by {84-196}\,$\times$ over the state-of-the-art.
    \item We implement Flash on CPU and further improve the performance by using multithreading, offloading data-independent computations to the offline phase, and employing lazy reduction in computing the proposed convolution.
    \item The system-wise optimized Flash improves the state-of-the-art PI by {17-46}\,$\times$ in end-to-end latency and {84-196}\,$\times$ in total communication cost. Moreover, Flash can process ImageNet models within a minute on CPU with the total communication less than 1\,GB.
\end{itemize}

%% file: Contents/2_background.tex
\section{Background}
\label{sec:background}
In this section, we describe our threat model (Section~\ref{subsec:threat}) and introduce the cryptographic primitives widely used for PI including homomorphic encryption (Section~\ref{subsec:he}), additive secret sharing (Section~\ref{subsec:ss}), garbled circuits (Section~\ref{subsec:gc}), and Beaver's triples (Section~\ref{subsec:bt}).
Existing PI protocols built upon the cryptographic primitives will be described in Section~\ref{subsec:existing}. 

\subsection{Threat Model}\label{subsec:threat}
Flash is a two-party privacy-preserving CNN inference system, 
where the cloud, or server, processes the client’s private data without gaining any information about individual data.
Simultaneously, Flash prevents any leakage of the cloud’s proprietary CNN model parameters such as weights to the client. 
However, both parties possess knowledge on hyperparameters and model architecture, e.g., the number and type of layers, and input/output dimensions of each layer~\cite{juvekar2018gazelle,mishra2020delphi,rathee2020cryptflow2}.

Following prior art~\cite{choi2022impala,gilad2016cryptonets,huang2022cheetah,juvekar2018gazelle,mishra2020delphi,rathee2020cryptflow2,roh2023hyena}, we assume that both parties, the cloud and the client, behave as semi-honest (i.e., honest but curious) adversaries. 
In other words, both parties follow the protocol honestly, but they may attempt to obtain extra information about the private data or model weights, which is not explicitly allowed by the protocol.

\begin{table}
\begin{threeparttable}
\setlength{\tabcolsep}{4pt}
\centering
\caption{Notations used in this paper.}
\label{table:notation}
{\small
\begin{tabularx}{\columnwidth}{p{0.9cm}X}
\toprule[1.5pt]
  & \textbf{\textit{Description}} \\
\midrule[0.2pt]
\midrule[0.2pt]
$n$ & Degree of plaintext and ciphertext polynomial + 1. \\
$p$ & Plaintext modulus. \\
$q$ & Ciphertext modulus. \\
$\Delta$ & Scale multiplied during encryption, defined as $\lfloor \frac{q}{p} \rfloor$. \\
${\mathbf{m}}$ & Message (data) vector with $n$ slots (i.e., dimension $n$).\\
$\mathbf{0}$ & $n$-slot message vector whose elements are all zeros.\\
$m(x)$ & Plaintext polynomial encoding a message vector $\mathbf{m}$.\\
$\llbracket\mathbf{m}\rrbracket$ & Ciphertext encrypting a message $\mathbf{m}$.\\ 
\bottomrule[1.5pt]
\end{tabularx}
}
\end{threeparttable}
\vspace{-1em}
\end{table}

\subsection{Homomorphic Encryption}\label{subsec:he}
Since the introduction of HE~\cite{gentry2009fully}, many encryption schemes have been proposed, such as BFV~\cite{fan2012somewhat}, BGV~\cite{brakerski2014leveled}, and CKKS~\cite{cheon2017homomorphic}, built upon the Ring Learning with Errors (RLWE) problem~\cite{lyubashevsky2010ideal}.
Typically a message vector $\mathbf{m}$ with fixed-point or integer elements is encoded using \textit{batch encoding} (or SIMD encoding)~\cite{smart2014fully} to a plaintext polynomial denoted as ${m(x)}$ within the polynomial ring $R_p=\mathbb{Z}_p[x]/(x^n+1)$ (i.e., $(n-1)$-th degree polynomial with integer coefficients in $(-p/2, p/2]$).
(Notations adopted in this paper are summarized in Table~\ref{table:notation}.)
HE schemes with batch encoding enable parallel arithmetic operations such as Single-Instruction-Multiple-Data (SIMD) addition and SIMD multiplication (i.e., element-wise addition/multiplication) between ciphertexts.
Leveraging the SIMD operations can improve the computation and communication efficiency by a factor of $n$ as batch encoding allows packing up to $n$ data in a plaintext\footnote{CKKS allows packing up to $n/2$ data in a plaintext~\cite{cheon2017homomorphic}.}~\cite{boemer2019ngraph, brutzkus2019low, gilad2016cryptonets, juvekar2018gazelle}.
However, linear operators in CNNs require computation between data in different slots, thereby necessitating slot rotation in a ciphertext, which is the most time-consuming operation as will be discussed shortly.
Flash employs BFV for data encryption but encodes a data vector in a different manner, which greatly reduces the slot rotation latency compared to batch encoding.


\ignore{In this paper, we use the BFV scheme because we can pack twice as many data as CKKS, thereby reducing the communication cost.}


\begin{table}
\begin{threeparttable}
\setlength{\tabcolsep}{4pt}
\centering
\caption{Comparison between public-key and private-key encryption for BFV. (Notation. (${p_0(x)}$, ${p_1(x)}$): public key, ${s(x)}$: secret key, ${e_0(x), e_1(x), u(x)}$: random polynomials, 
and (${c_0(x), c_1(x)}$): ciphertext encrypting $m(x)$. $[\cdot]_q$ denotes applying $\mod q$ to all the coefficients.)}
\label{table:enc_method}
{\small
\begin{tabular}{@{}c@{}cc@{}rcc}
\toprule[1.5pt]
& \textbf{\textit{Public-key encryption}} & \textbf{\textit{Private-key encryption}} \\
\midrule[0.2pt]
\midrule[0.2pt]
\multirow{2}{*}{${c_0(x)}$} & \multicolumn{1}{c}{\multirow{2}{*}[-0.ex]{\parbox{2.8cm}{\centering {$[{p_0(x) u(x)}+{e_0(x)}$ \\$+ \Delta m(x)]_q$}}}} & \multicolumn{1}{c}{\multirow{2}{*}[-0.ex]{\parbox{2.8cm}{\centering{$[-({a(x)s(x)}+{e_0(x)})$\\$+\Delta {m(x)}]_q$}}}} \\ [0.5mm]
&&&\\[0.5mm]
${c_1(x)}$ & $[{p_1(x) u(x)} + {e_1(x)}]_q$ & $a(x)$\\ [0.5mm]
{Enc. latency} & 366.7\,$\micro\textrm{s}$ & 248.1\,$\micro\textrm{s}$ \\ [0.5mm]
Noise budget & 31 bits & 37 bits  \\ [0.5mm]
{Online enc. latency\footnotemark[1]} & 26.3\,$\micro\textrm{s}$ & 26.3\,$\micro\textrm{s}$ \\ [0.mm]
\bottomrule[1.5pt]
\end{tabular}
}
\begin{tablenotes}
\footnotesize
\item [1] This will be explained in Section~\ref{sec:offline}.
\end{tablenotes}
\end{threeparttable}
\vspace{-1em}
\end{table}

With batch encoding, an $n$-dimensional vector $\mathbf{m}$, where the $i$-th entry is denoted by $\mathbf{m}[i]$, is encoded as a polynomial plaintext ${m(x)}$.
This is then encrypted using either a public or secret key, resulting in a ciphertext $\llbracket \mathbf{m}\rrbracket \in R_q^2$, comprised of two polynomials. 
Table~\ref{table:enc_method} details both the public-key and private-key encryption.
Flash uses private-key encryption only because 1) the cloud does not need to encrypt any data and 2) private-key encryption has lower latency and a larger noise budget than public-key encryption with the identical encryption parameters.

The following operations are available in HE with batch encoding: homomorphic addition of two ciphertext (\textbf{HAdd}), multiplication of a plaintext with a ciphertext (\textbf{PMult}), multiplication of a constant with a ciphertext (\textbf{CMult}), and cyclic slot rotation of a ciphertext (\textbf{HRot})\footnote{Multiplication between ciphertexts, which Flash does not use, is omitted here.}. Each operation is performed as follows:
(let $\llbracket \mathbf{m_0}\rrbracket=({c_0(x)}, {c_1(x)})$ and $\llbracket \mathbf{m_1}\rrbracket=({c'_0(x)}, {c'_1(x)})$\footnote{We assume that number theoretic transform (NTT) has been applied to both $\llbracket \mathbf{m_0}\rrbracket$ and $\llbracket \mathbf{m_1}\rrbracket$ so they are in the evaluation space (or in NTT domain)~\cite{brakerski2014leveled}.}.)
\ignore{, which encrypt the plaintexts ${m_0(x)}$ and ${m_1(x)}$, respectively encoding vectors $\mathbf{m_0}$ and $\mathbf{m_1}$.}
\begin{align}
\textbf{HAdd}(\llbracket \mathbf{m_0}\rrbracket,\llbracket \mathbf{m_1}\rrbracket)&{=}\llbracket \mathbf{m_0}\text{+}\mathbf{m_1}\rrbracket \nonumber \\
&{=}({c_0(x)}+{c'_0(x)}, {c_1(x)}+{c'_1(x)}) \nonumber \\
\textbf{PMult}(\llbracket \mathbf{m_0}\rrbracket,{m_1(x)})&{=}\llbracket \mathbf{m_0 m_1}\rrbracket \nonumber \\
&{=}({c_0(x)}\times {m_1(x)}, {c_1(x)}\times {m_1(x)}) \nonumber \\
\textbf{CMult}(\llbracket \mathbf{m_0}\rrbracket,a) &{=}\llbracket a\mathbf{m_0}\rrbracket \nonumber \\
&{=}(c_0(x)*a, c_1(x)*a) \nonumber \\
\textbf{HRot}(\llbracket \mathbf{m_0}\rrbracket,step)&{=}\llbracket\langle \mathbf{m_0}\rangle_{step}\rrbracket \nonumber
\end{align}
where $+$, $\times$ and $*$ represent coefficient-wise addition, coefficient-wise multiplication, and multiplication of a constant with all coefficients, respectively.
Noise in output ciphertext increases additively for \textbf{HAdd} and multiplicatively by a factor of around $\sqrt{n}p/2$ for \textbf{PMult}~\cite{juvekar2018gazelle} and a factor of the multiplying constant $a$ for \textbf{CMult}.
$\langle \mathbf{m_0}\rangle_{step}$ denotes the left-cyclic slot shift of $\mathbf{m_0}$ by $step$.
For instance, when $\mathbf{m_0}=(\mathbf{m_0}[0], \mathbf{m_0}[1], \dots, \mathbf{m_0}[n-1])$, $\langle \mathbf{m_0}\rangle_{step}$ returns $(\mathbf{m_0}[step], \dots, \mathbf{m_0}[n-1], \mathbf{m_0}[0], \dots, \mathbf{m_0}[step-1])$.

\textbf{HRot} procedure consists of two main stages:
1) Applying inverse number theoretic transform (INTT) to the ciphertext, which makes $c_0(x)$ and $c_1(x)$ in the coefficient space, 
then decomposing each coefficient of the ciphertext polynomials, followed by applying NTT to each decomposed polynomial to place them back in the evaluation space, as referenced in~\cite{brakerski2014leveled,fan2012somewhat}, and 2) automorphism and key-switching. 
Decomposition is used to segment polynomials into multiple components with smaller valued coefficients, which prevents substantial noise growth during key-switching.
For instance, when $2^T$ is chosen as the decomposition base, 
a single ciphertext $\llbracket \mathbf{m_0}\rrbracket$ with modulus $q$ is divided into $l$ ciphertexts $\llbracket \mathbf{m_0}\rrbracket^{(l)} = ({c_0(x)}, {c_1(x)}^{(l)})$ with $l \in [1, \lceil \log_{2^{T}}q\rceil]$.
Subsequently, these $l$ decomposed ciphertexts undergo NTT operations.

Then, applying automorphism to $l$ ciphertexts $\llbracket \mathbf{m_0}\rrbracket^{(l)}$ returns $\llbracket \mathbf{m_0}\rrbracket^{'(l)}$ = $({c_0(x^{3^{step}})}, {c_1(x^{3^{step}})^{(l)}})$. 
Automorphism basically relocates the coefficients of the polynomials from index $i$ to $i\cdot 3^{step}\ \text{mod}\ n$ for all $i \in [0,n-1]$.
Finally, key-switching is performed on $\llbracket \mathbf{m_0}\rrbracket^{'(l)}$ as the last step for \textbf{HRot}.
To this end, key-switching keys, denoted as ($swk_0^{step}(x)^{(l)}$, $swk_1^{step}(x)^{(l)}$) for each $step$, should be provided by the client and stored in the cloud.
Key switching is then performed as follows:
\begin{gather}
\label{eq:keySwitching}
c_0'(x)=c_0(x^{3^{step}}) + \sum_{i=1}^{l} swk_0^{step}(x)^{(i)}\times c_1(x^{3^{step}})^{(i)} \nonumber \\
c_1'(x)=\sum_{i=1}^{l} swk_1^{step}(x)^{(i)}\times c_1(x^{3^{step}})^{(i)} \nonumber 
\end{gather}
and the ciphertext $(c_0'(x), c_1'(x))$ is returned by $\textbf{HRot}(\llbracket \mathbf{m_0}\rrbracket,step)$.

\begin{figure}[t]
\centering
\includegraphics[width=0.8\columnwidth]{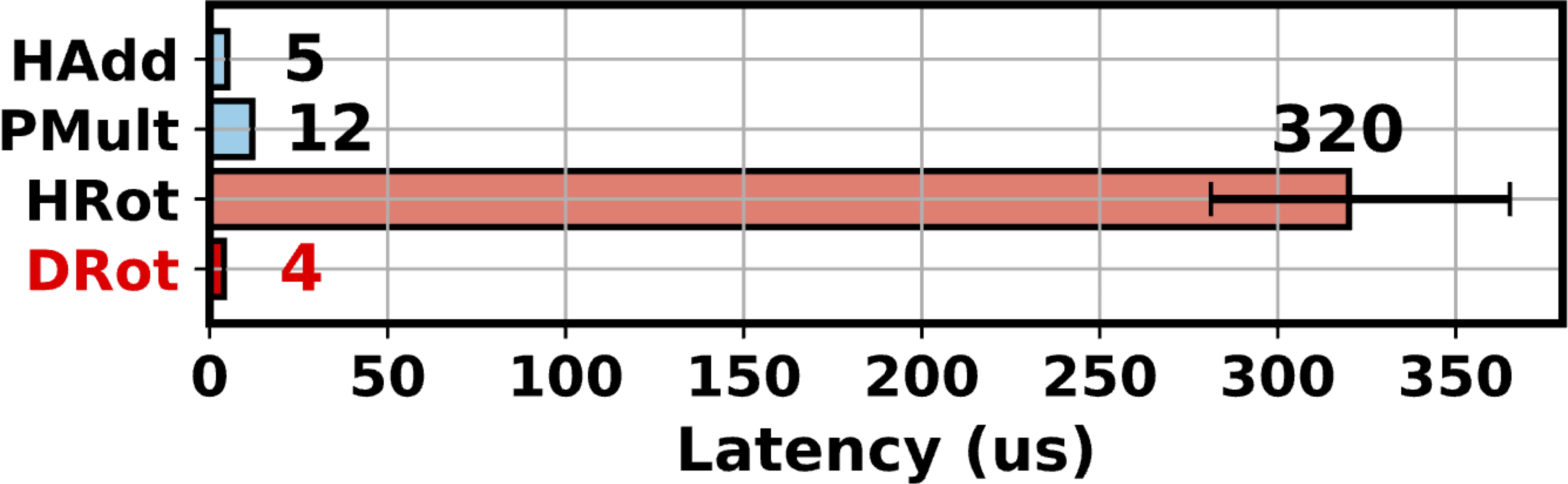}
\vspace{-0.5em}
\caption{Latency comparison between HE operations. Encryption parameters are chosen to compute convolutions in VGG-16 for ImageNet, and \textbf{\textbf{HRot}} latency varies with decomposition base.}
\label{fig:operation}
\vspace{-0.5em}
\end{figure}

Among HE operations described so far, it is obvious that $\textbf{HRot}$ is the most time-consuming operation since it requires a large number of polynomial multiplication and NTT.
(Runtime of each HE operation is shown in Fig.~\ref{fig:operation}.)
Decreasing the amount of computation by increasing the decomposition base to reduce $l$, can lessen \textbf{HRot}'s latency, but it results in larger noise growth, and computation will be wrong if the noise in the ciphertext grows beyond the noise budget.
Therefore, determining the optimal decomposition base balancing latency and noise is critical.
As illustrated in Fig.~\ref{fig:operation}, runtime of \textbf{HRot} varies depending on the decomposition base, which allows correctly computing convolution layers in VGG-16 for ImageNet.

\subsection{Additve Secret Sharing}\label{subsec:ss}
Additive secret sharing (SS)~\cite{blakley1979safeguarding,adi1979share,damgaard2012multiparty} divides a private value $x$ among two or more parties so that no single party can deduce $x$ from their share. 
In the 2PC setting, one party chooses a random value $r$ and provides it to the other party, so the share of each party becomes $[x]_1=x-r$ and $[x]_2=r$, respectively.
The original value $x$ remains perfectly secret to the other party unless both shares are combined: $x=[x]_1+[x]_2$.

\subsection{Garbled Circuits}\label{subsec:gc}
Garbled circuit~\cite{yao1986generate} is a two-party protocol where two parties---a garbler and an evaluator---compute the output $z$ of a boolean circuit $C$ using their private inputs $x$ and $y$ without revealing these inputs to each other.

A garbler begins with “garbling” circuit $C$ into a garbled circuit (GC) $\tilde{C}$.
For each input wire of every gate, a pair of random labels is assigned, corresponding to 0 and 1. 
The garbler then produces an encrypted truth table, known as the garbled table. 
This table maps the output labels to the corresponding gate input labels for each gate in the circuit.
Following this, the garbler transmits GC and the input labels corresponding to its private input $x$ to the evaluator. 
The evaluator, using Oblivious Transfer (OT)~\cite{ishai2003extending}, acquires the input labels corresponding to the evaluator's private input $y$.
OT ensures that the garbler remains unaware of $y$.
The evaluator evaluates each gate in GC sequentially to obtain the result and finally sends the output labels back to the garbler.
The garbler can derive the boolean circuit output $z$ from the received output labels.

In 2PC-based PI, the server and client act as a garbler and evaluator to compute ReLU by representing it using boolean circuits~\cite{patra2020blaze,riazi2018chameleon,riazi2019xonn,mohassel2018aby3}. 
However, challenges arise from the time-intensive garbling and evaluating processes, the large GC size, and the significant communication costs during the GC transmission and OT protocol.

\subsection{Beaver's Triples}\label{subsec:bt}

While SS in Section~\ref{subsec:ss} allows computing the summation $x+y$ easily from the private values $x$ and $y$ of each party, 
computing the product $xy$ using SS is quite involved, necessitating multiple rounds of communication between the parties.
Beaver’s triples (BT)~\cite{beaver1995precomputing}, comprising random values $a$ and $b$ and their product $ab$, assist in this multiplication. 
The process has two phases:
\begin{itemize}
\item \textbf{Generation of triples}: multiplication triples are generated in advance. Party one gets shares $[a]_1$, $[b]_1$, $[ab]_1$; party two gets $[a]_2$, $[b]_2$, $[ab]_2$. 
\item \textbf{Multiplication}: the values $x$ and $y$ are secret-shared. Using the pre-generated triples, the multiplication is executed, resulting in the product shares $[xy]_1$ for party one and $[xy]_2$ for party two.
\end{itemize}
During the entire process, nothing about $x$, $y$, or $xy$ is leaked.
Note that SS, GC and BT cannot be reused for security and they should be newly generated for every operation. 

\subsection{Existing PI Protocols}\label{subsec:existing}


Since CNNs are divided into linear and nonlinear layers, how to efficiently implement the linear and nonlinear operators in CNNs using the cryptographic primitives has been actively studied for PI. 

PIs processing both linear and nonlinear operators using HE only are one category, 
which can be again divided into either leveled-HE (LHE)-based PIs that does not use bootstrapping or fully HE (FHE)-based ones with bootstrapping~\cite{al2023demystifying}.
LHE-based PIs~\cite{gilad2016cryptonets,hesamifard2017cryptodl,brutzkus2019low} choose the optimal encryption parameters to minimize the computational overhead, while allowing correct computation without noise overflow, once the model is decided.
The downside of this scheme is that PI for deep CNNs suffers from severe latency degradation
because large multiplicative depth enforces choosing excessively large encryption parameters,
causing enormous computational overhead.
Hence, this technique can be applied to shallow CNNs with low inference accuracy only.

On the other hand, FHE-based PIs can process deep CNNs such as VGG or ResNet models by employing bootstrapping that reduces noise in the ciphertext to support large multiplicative depth.
However, bootstrapping consumes a significant runtime and needs to be performed after every nonlinear activation function periodically,
and many recent works have studied to accelerate bootstrapping in algorithm level~\cite{lee2021high,bossuat2021efficient}. 
However, even with those efforts, latency of the state-of-the-art FHE-based PI on CPU~\cite{lee2021high} with batch size of 1 takes more than 62 minutes for ResNet-32 with CIFAR-10.
(Inference for ResNet-32 with CIFAR-100 takes more than 65 minutes.)
Accelerating bootstrapping using GPUs~\cite{jung2021over,kim2023hyphen} or custom accelerators~\cite{kim2022bts,kim2022ark,samardzic2021f1} can be an option, 
but hardware acceleration is orthogonal to the contributions of this paper,
where Flash is optimized in algorithm level and implemented on CPU.

Instead of using HE only, hybrid protocols utilizing multiple cryptographic primitives to optimize the PI performance has been adopted in other prior works~\cite{juvekar2018gazelle,rathee2020cryptflow2,reagen2021cheetah,roh2023hyena}. 
In case of linear operators, for instance, Gazelle~\cite{juvekar2018gazelle} proposes packed convolution using HE, where multiple input channels are packed into a single ciphertext and SIMD operations are exploited, to reduce both computation/communication overhead.
For handling ReLU operations, many works~\cite{liu2017oblivious,mohassel2018aby3,mohassel2017secureml,yao1982protocols} including Gazelle utilize GC. 
While GC enables to process ReLU without any loss of accuracy, thereby achieving high inference accuracy, it incurs substantial communication costs, ultimately leading to increased latency.
To reduce the overhead associated with ReLU, prior art such as \cite{demmler2015aby,mohassel2018aby3,rathee2020cryptflow2,wagh2019securenn} have developed faster algorithms for ReLU,
while \cite{jha2021deepreduce,cho2022selective,cho2022sphynx,kundu2023learning} have modified model architectures to use fewer ReLUs, incurring an accuracy drop.


\begin{figure*}[t]
\centering
\includegraphics[width=1.8\columnwidth]{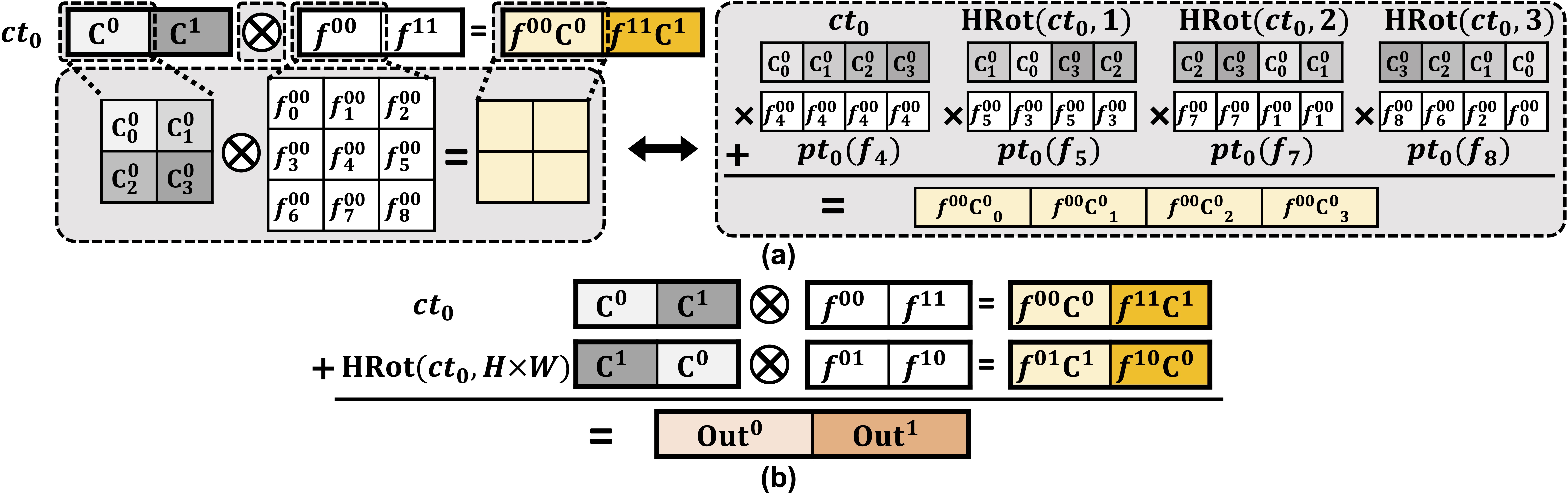}
\vspace{-0.5em}
\caption{Conventional multi-channel convolution. Note that superscripts indicate the order of input channels: (a) single convolution process with two channels packed, and (b) channel-rotation to add partial sums.}
\label{fig:conv_conventional}
\vspace{-0.5em}
\end{figure*}

Delphi~\cite{mishra2020delphi} divides the whole inference process into an offline and an online phase. 
This allows for intensive cryptographic computations, such as creating a large amount of secret shares using HE, to be pre-executed during the offline phase. 
Then, in the online phase, when actual inputs are received, convolutions are efficiently processed using SS.
For handling ReLU, the protocol sacrifices the inference accuracy by either partially or fully replacing ReLUs with polynomial activation functions. 
These functions are processed using BT, which offer a more cost-effective solution in terms of both computation and communication when compared to GC~\cite{park2022aespa}.
If accuracy is a priority, Delphi can be implemented using an all-ReLU (GC) structure, similar to the implementation in \cite{park2022aespa,garimella2023characterizing}.
However, as pointed out in \cite{garimella2023characterizing}, if the computation time and communication cost of the offline phase is extended, it can subsequently affect the latency in the online phase.
Flash solves this issue by proposing a new training algorithm that can achieve high inference accuracy with deep CNNs with polynomial activation functions and by using a novel 2PC protocol that can evaluate a polynomial without any offline communication. 

Some PI protocols using HE for linear operators, such as \cite{xu2023falcon,huang2022cheetah}, adopt different encoding schemes, not batch encoding, to avoid costly slot rotation (\textbf{HRot}).
However, the convolution algorithms proposed in \cite{xu2023falcon,huang2022cheetah} without slot rotation leads to significant under-utilization of slots in the output ciphertexts and may result in large communication overhead.
The optimization technique presented in \cite{xu2023falcon} to reduce the communication cost is only applicable to depthwise convolution. 
Moreover, they also require offline communication to reduce online latency for computing convolution.
On the contrary, Flash proposes a convolution that does not require any offline communication and that has low latency and communication cost as will be described in Section~\ref{sec:conv}. 






%% file: Contents/3_conv.tex
\section{Convolution with Direct Encoding}
\label{sec:conv}

\ignore{
\begin{figure}[t]
\centering
\includegraphics[width=0.9\columnwidth]{Figure/encoding.jpg}
\caption{Proposed encoding and rotation compared with conventional method. Details of the proposed methods are elaborated in Algorithm~\ref{alg:direct_encoding} and~\ref{alg:direct_rotation}.}
\label{fig:encoding}
\vspace{-1em}
\end{figure}
}

In this section, we propose a new convolution algorithm along with a novel data vector encoding scheme to improve the latency of convolution, which constitutes a significant portion of the PI end-to-end latency~\cite{roh2023hyena}.
To quantify the performance improvement of the proposed convolution, we compare our results with the state-of-the-art convolutions~\cite{mishra2020delphi, juvekar2018gazelle, brutzkus2019low}.

\subsection{Conventional Convolution} \label{sec:3.0}
The client encodes an input feature map in row-major order to a plaintext polynomial using batch encoding, which is then encrypted into a ciphertext~\cite{dathathri2019chet, halevi2014algorithms, kim2022secure}.
When the input feature map size is large (e.g., TinyImageNet or ImageNet cases), 
input should be divided into $n$-dimensional vectors, each of which is encrypted into a ciphertext, to minimize the required number of ciphertexts and reduce the communication cost.


Consider, for example, a three-dimensional tensor with channel $C$, height $H$, and width $W$.
To fully utilize the slots in a ciphertext, $C\times H\times W$ input elements should be packed into $\lceil \frac{C\times H\times W}{n} \rceil$ ciphertexts.
Depending on the magnitude of $n$ and $H\times W$, each ciphertext may contain multiple channels of input feature maps, or multiple ciphertexts may be required to encrypt a single channel (if $n < H\times W$).

To perform the conventional convolution, the cloud encodes the kernel elements to the plaintexts with batch encoding. 
Each element in the plaintext is multiplied to the correct input feature map element to produce the partial sums using \textbf{PMult}.
Consider, for instance, the case depicted in Fig.~\ref{fig:conv_conventional}, which involves 2$\times$2$\times$2 input and output channels, associated with 3$\times$3 filters.
In this case, the inputs from channel 0 and 1 (\textbf{C$^0$} and \textbf{C$^1$}) are packed together into a single ciphertext \textbf{$\bm{ct}{_\text{\textbf{0}}}$}. 
Subsequently, the kernels, \textbf{$f$$^{\text{\textbf{out ch., in ch.}}}$}, associated with each input channel for the different output channels  (\textbf{$f{^\text{{00}}}$} and \textbf{${f^\text{{11}}}$}) are also packed together into plaintexts {$\bm{pt}{_{\textbf{0}}}$}'s. 
Then, for convolution, as shown in Fig.~\ref{fig:conv_conventional}(a), the ciphertext undergoes four rotations (\textbf{\textbf{HRot}}), and \textbf{$\bm{pt}{_\text{\textbf{0}}}(\cdot)$} packing appropriate kernel elements is multiplied (\textbf{PMult}) with the rotated ciphertext.
The products of these multiplications are summed together to yield a ciphertext that contains the intermediate partial sums for the output channels.
Finally, these partial sums undergo rotations to align their respective positions and are then added together to obtain the convolution output as shown in Fig.~\ref{fig:conv_conventional}(b). 
Note that the output ciphertext \textbf{$\llbracket$Out$^0\vert \vert$Out$^1\rrbracket$} is obtained with the output channels packed, which helps lower the communication cost in PI by reducing the number of ciphertexts to be transmitted.

\begin{figure*}[t]
\centering
\includegraphics[width=1.8\columnwidth]{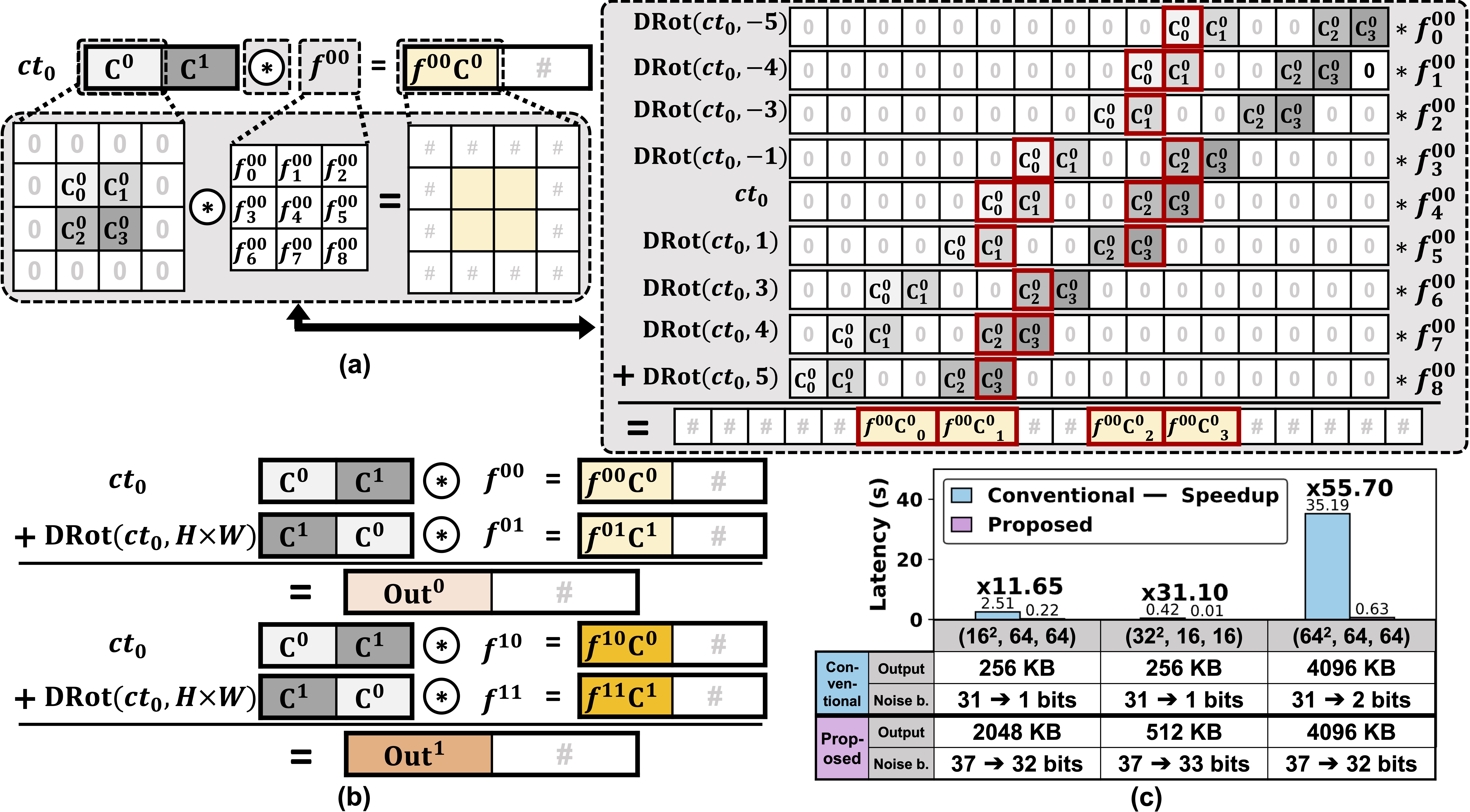}
\vspace{-0.5em}
\caption{Proposed multi-channel convolution. Superscripts indicate the order of input channels and \# denotes slots occupied with dummy data: (a) single-channel convolution process with a ciphertext packing two input channels, (b) channel-rotation to add partial sums, and (c) comparison of latency, output ciphertext size, and remaining noise budget between conventional and proposed convolution across various parameter sets ($H\times W$, $c_i$, $c_o$) with 3$\times$3 kernels.}
\label{fig:conv_proposed}
\vspace{-0.5em}
\end{figure*}

\subsection{Proposed Slot Rotation over Encrypted Data} \label{sec:3.1}

Conventional convolution with batch encoding involves \textbf{\textbf{HAdd}}, \textbf{\textbf{PMult}}, and \textbf{\textbf{\textbf{HRot}}}~\cite{brutzkus2019low,juvekar2018gazelle,reagen2021cheetah}. 
Among these, \textbf{\textbf{HRot}} introduces the largest latency penalty, mainly because of the compute-intensive key switching operation.
As plotted in Fig.~\ref{fig:operation}, for the encryption parameters to compute convolution layers in VGG-16 for ImageNet, \textbf{\textbf{HRot}} is 23-30\,$\times$ slower than \textbf{PMult}.

To accelerate convolution by orders of magnitude, we propose a novel encoding scheme, called direct encoding, that enables efficient rotation operation over encrypted data, which we call direct rotation \textbf{DRot}. 
Note that the proposed convolution is implemented such that SIMD multiplication (element-wise multiplication between two different vectors) is not required, so batch encoding, which allows both SIMD addition and SIMD multiplication, is avoided in Flash.
Direct encoding is different from batch encoding in that data are directly placed onto the coefficients of a plaintext.
That is, a data vector $\mathbf{m}$ of length $n$ is mapped to a polynomial plaintext $m(x)$ as follows:
\begin{equation}
\label{eq:de}
\mathbf{m} \mapsto m(x) = \sum_{i=0}^{n-1} \mathbf{m}[i]\cdot x^i
\end{equation}
\ignore{This approach greatly simplifies the rotation operation, which is simply adjusting the positions of the coefficients of the ciphertext \fixme{like Algorithm~\ref{alg:direct_rotation}}, thereby eliminating time-consuming key-switching operations.}

\ignore{
\fixme{This encoding greately simplifies the rotation operation, which is simply adjusting the positions of the coefficients of the ciphertext.
To operate $\textbf{DRot}(\llbracket m\rrbracket,step){=}\llbracket\langle m\rangle_{step}\rrbracket$, consider the following: Let $\llbracket m\rrbracket=(c_0, c_1)$, where $\llbracket\langle m\rangle_{step}\rrbracket=(c'_0, c'_1)$, $q=\text{ciphertext modulus}$,
$P_i = {i+step > n-1}$, and 
$Q_i = {i+step < 0}$.
\begin{equation}
\forall j \in \{0, 1\}, c'_j \mapsto
\sum_{i\in N} \left\{ 
  \begin{array}{l}
    (q-c_j[i+step-n])\cdot x^i \text{  if } P_i\\
    (q-c_j[i+step+n])\cdot x^i \text{  if } Q_i\\
    c_j[i+step]\cdot x^i \text{  otherwise}
  \end{array} 
\right.
\end{equation}
}
}

This encoding can significantly simplify the rotation operation, which simply rearranges the positions of the ciphertext coefficients.
Specifically, $\textbf{DRot}(\llbracket \mathbf{m}\rrbracket,step){=}\llbracket\llangle \mathbf{m}\rrangle_{step}\rrbracket$ is performed as follows: (let $\llbracket \mathbf{m}\rrbracket=(c_0(x), c_1(x))$ and $\llbracket\llangle \mathbf{m}\rrangle_{step}\rrbracket=(c'_0(x), c'_1(x))$)\footnote{Proof on why \eqref{eq:drot} performs rotation over encrypted data with direct encoding is provided in Appendix~\ref{subsec:proof}.}
\begin{equation}
\label{eq:drot}
\mathrm{For}\, j \in \{0, 1\}, \quad c'_j (x) = c_j (x)\cdot x^{-step} \mod x^n+1
\end{equation}
$\llangle \mathbf{m}\rrangle_{step}$ denotes left-cyclic slot shift of $\mathbf{m_0}$ by $step$ with the sign inverted upon wraparound.
For instance, when $\mathbf{m}=(\mathbf{m}[0], \mathbf{m}[1],$ $\dots$$, \mathbf{m}[n-1])$, $\llangle \mathbf{m}\rrangle_{step}$ returns $(\mathbf{m}[step],$ $ \dots$$, \mathbf{m}[n-1], \mathbf{-m}[0]$, $\dots$$, \mathbf{-m}[step-1])$\footnote{This sign inversion does not cause wrong computation in the proposed convolution thanks to padded zeros as illustrated in Fig.~\ref{fig:conv_proposed}.}.

\textbf{DRot} provides several advantages over \textbf{HRot}.
First, as evident from \eqref{eq:drot}, by eliminating time-consuming key-switching operations, runtime of \textbf{DRot} is lower than \textbf{HRot} by orders of magnitude, even less than that of \textbf{HAdd} as illustrated in Fig.~\ref{fig:operation}.
Second, \textbf{DRot} does not increase any noise in the ciphertext due to the absence of key switching. (Proof is given in Appendix~\ref{subsec:proof}.)
Consequently, it allows extremely low latency for computing the convolution layers with a large number of channels, which are common in deep CNN models.
Finally, \textbf{DRot} does not require any switching keys from the client, which helps save a large amount of storage for PI as will be explained in Section~\ref{subsec:proposed_conv}.

Note that what we propose in this work is the encoding scheme and its associated slot-rotation operation, not the encryption scheme. 
The data privacy in Flash is guaranteed since the encoded data is encrypted using BFV, which is IND-CPA secure (i.e., BFV ciphertexts of any two messages \textbf{u} and \textbf{u'} are computationally indistinguishable). 

\subsection{Proposed Convolution with DRot}
\label{subsec:proposed_conv}

Flash takes advantage of \textbf{DRot} with direct encoding to compute convolution with low latency,
and the proposed convolution is illustrated in Fig.~\ref{fig:conv_proposed}.
Since Flash uses direct encoding, not batch encoding, slot-wise multiplication using \textbf{PMult} is not supported, but multiplying an identical value to all the slots (i.e., multiplying a constant value to all the encoded data simultaneously) using \textbf{CMult} can be performed.
Hence, the proposed convolution is implemented using \textbf{CMult} and \textbf{DRot} as follows\footnote{At a high-level implementation, the proposed convolution is similar to the padded convolution in \cite{juvekar2018gazelle} that uses batch encoding. Thanks to the proposed \textbf{DRot}, the convolution runtime of Flash becomes up to 94\,$\times$ faster when the input is large.}.
The cloud receives the ciphertexts that encrypt zero-padded inputs from the client.
Single-channel convolution is computed using \textbf{DRot} and \textbf{CMult} as shown in Fig.~\ref{fig:conv_proposed}(a). 
Then, these partial sums generated from multiple input channels are aligned using \textbf{DRot} and added to obtain the final multi-channel convolution outputs as depicted in Fig.~\ref{fig:conv_proposed}(b).
This procedure is detailed in Algorithm~\ref{alg:conv}.
Compared to prior art, the proposed convolution has the following features.



First, unlike \cite{juvekar2018gazelle,reagen2021cheetah,choi2022impala}, since convolution is implemented with \textbf{CMult}, not \textbf{PMult}, the cloud does not need to store the kernel weights for the convolution layers in the form of plaintexts~\cite{roh2023hyena}.
In addition, thanks to using \textbf{DRot}, since the switching keys are not required for slot rotation, Flash can substantially reduce the storage requirements for PI.
As summarized in Table~\ref{table:server_storage}, this feature result in significant cloud storage savings with a total reduction of 1962\,$\times$ in the case of VGG-16 for ImageNet\footnote{Conventional convolution can be implemented using \textbf{CMult} to save storage at the cost of increased runtime. Here we assume runtime-optimized implementation.}.

\begin{table}
\begin{threeparttable}
\setlength{\tabcolsep}{4pt}
\centering
\caption{Total server storage usage for all convolution layers in VGG-16 for ImageNet: conventional~\cite{brutzkus2019low,juvekar2018gazelle,reagen2021cheetah} vs. proposed method.}
\label{table:server_storage}
{\small
\begin{tabular*}{\columnwidth}{@{\extracolsep{\fill}}ccc@{}rcc}
\toprule[1.5pt]
\multicolumn{1}{c}{\multirow{2}{*}[-0.65ex]{\parbox{1.5cm}{\centering \textbf{\textit{Storage \\ component}}}}} & \multicolumn{2}{c}{\textbf{\textit{Conventional}}} &
& \multicolumn{2}{c}{\textbf{\textit{Proposed}}} \\
\cmidrule{2-3}\cmidrule{5-6}
& {\textbf{Size}} & {\textbf{Related op.}} & & {\textbf{Size}} & {\textbf{Related op.}}\\
\midrule[0.2pt]
\midrule[0.2pt]
Weights & 215\,GB & \textbf{PMult} & & 0.11\,GB & \textbf{CMult} \\ [0.mm]
Switching keys & 12\,MB & \textbf{\textbf{HRot}} & & 0\,MB & \textbf{DRot} \\ [0.mm]
\textbf{Total saving} & \multicolumn{2}{c}{\textbf{1$\times$}} & & \multicolumn{2}{c}{\textbf{1962$\times$}}\\ [-0.2mm]
\bottomrule[1.5pt]
\end{tabular*}
}
\end{threeparttable}
\end{table}

Second, since \textbf{DRot} runs much faster than \textbf{\textbf{HRot}}, a significant speedup in computing convolution can be achieved.
For several convolution layers, this leads to 12-56\,$\times$ speedup (see Fig.~\ref{fig:conv_proposed}(c)). 
Here, $c_i$ and $c_o$ represent the number of input and output channels, respectively.
As mentioned in Section \ref{sec:3.1}, since \textbf{DRot} does not add any noise, the proposed convolution adds much less noise during computation than the conventional method as shown in Fig.~\ref{fig:conv_proposed}(c).
In the conventional approach, 
the remaining noise budget significantly drops from 31 bits to 1 bit after convolution when the encryption parameters are optimized for lowest latency.
However, for the same convolution layers, the proposed method reduces the noise budget from 37 to 32 bits, while latency is improved by more than 10-90\,$\times$\footnote{More comparison on convolution runtime is provided in Appendix~\ref{subsec:conv}.}.
This allows handling convolution layers with higher computational loads without performance degradation.

One drawback of the proposed convolution is that 
it does not support output packing and separate ciphertexts are required for each output channel as illustrated in Fig.~\ref{fig:conv_proposed}(b). 
Thus, the output ciphertext size increases compared to the conventional method,
and Fig.~\ref{fig:conv_proposed}(c) shows the comparison for several convolution layers.
This will slightly increase the communication cost, but the performance benefit we can obtain from the proposed convolution is much bigger, which will be validated in Section~\ref{sec:results}.

\ignore{\textit{However, the size of the ciphertext generated in this process is only the largest communication cost when building an end-to-end network, including all activation layers and the entire process!}}


\begin{algorithm}[t]
\caption{Proposed convolution}
\label{alg:conv}
\DontPrintSemicolon
\SetArgSty{textnormal}
\KwIn{
Ciphertext vector $x$, filters $\in \mathbb{Z}^{c_o\times c_i\times {f_w}^2}$
}
\KwOut{Ciphertext vector $y$}

    \text{Let $c_n$=number of channel packing} \par
    \For {$i$ \text{  in length of $c_o$}} {
        \text{Initialize} \text{$\llbracket\mathbf{partial\ sum}\rrbracket$} to $\llbracket \mathbf{0}\rrbracket$\par
        \For {$j$ \text{  in length of $c_i$}} {

            \text{Let} $\llbracket \mathbf{c}\rrbracket$ = $x[j/c_n]$\par
            \text{Initialize} \text{$\llbracket\mathbf{t}\rrbracket$} to $\llbracket \mathbf{0}\rrbracket$\par
            \For {$k$ \text{  in length of ${f_w}^2$}} {
                \text{$\llbracket\mathbf{t}\rrbracket$} += \FuncSty{DRot}($\llbracket \mathbf{c}\rrbracket$, $k$)$*\text{filters}[i][j][k]$\;
            }
            \text{$\llbracket\mathbf{partial\ sum}\rrbracket$} += \FuncSty{DRot}(\text{$\llbracket\mathbf{t}\rrbracket$}, $H\times W\times (j\text{ mod }c_n)$)\;
        }
        $y[i]=$\text{$\llbracket\mathbf{partial\ sum}\rrbracket$}\;
    }
\end{algorithm}

%% file: Contents/4_training.tex
\section{Training with Polynomial Activation}
\label{sec:training}

Although the proposed convolution in Section~\ref{sec:conv} remarkably reduces the computational overhead in convolution for deep CNNs, 
accelerating convolution only is not sufficient since evaluating nonlinear activation functions remains a huge bottleneck in PI.
ReLU is the most widely used activation function for many deep CNNs with high accuracy.
However, securely evaluating ReLU has relied on GC~\cite{yao1986generate}, which incurs huge communication and storage overhead due to the required bitwise operations.
Despite many algorithmic optimizations~\cite{ishai2003extending,bellare2013efficient,zahur2015two}, single ReLU evaluation requires more than 2\,KB of data communication~\cite{choi2022impala}, and the amount of communication grows as the number of bits for the activations increases.
This implies that several GB of data communication is still required \textit{per inference} to run the ResNet or VGG models for processing CIFAR-100 and tens of GB for TinyImageNet.

\begin{figure}[t]
\centering
\includegraphics[width=0.7\columnwidth]{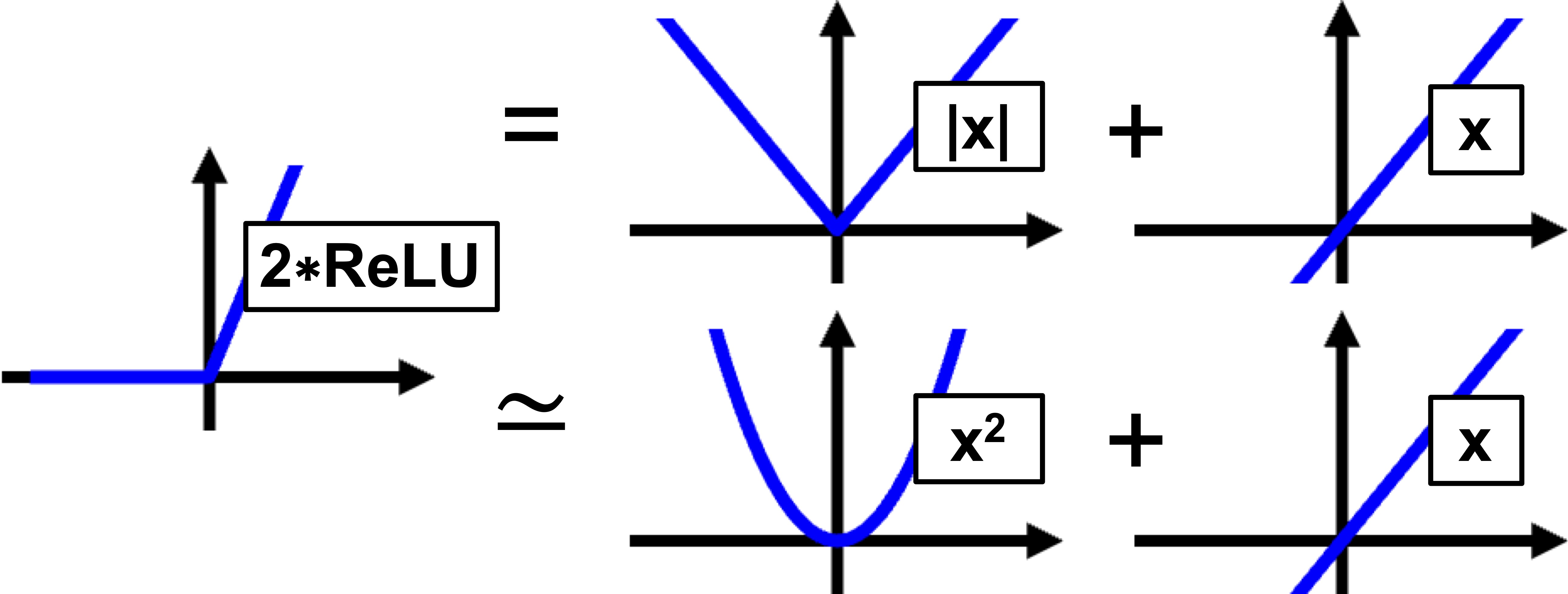}
\vspace{-0.5em}
\caption{Second-order polynomial approximation of ReLU.}
\label{fig:relu}
\vspace{-1em}
\end{figure}

\begin{figure*}[t]
\centering
\includegraphics[width=1.7\columnwidth]{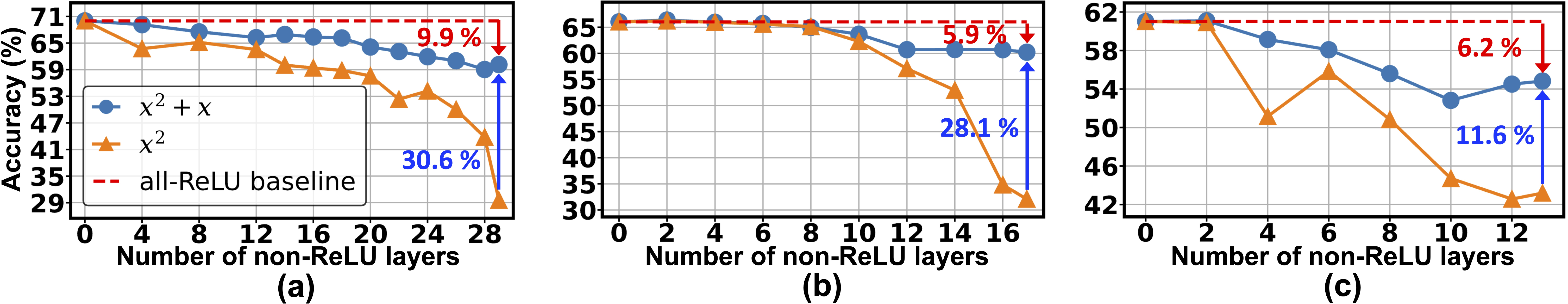}
\vspace{-0.8em}
\caption{Accuracy with replacement of ReLU: (a) ResNet-32 on CIFAR-100, (b) ResNet-18, and (c) VGG-16 on TinyImageNet.}
\label{fig:accuracy}
\vspace{-1em}
\end{figure*}


\begin{table}
\setlength{\tabcolsep}{4pt}
\centering
\caption{Test accuracy (\%) comparison for CIFAR-10/100 (C10/100) and TinyImageNet (Tiny) across multiple networks: Base. (all ReLU), \ignore{poly. (all ReLU replaced with polynomial approximations)}, Prior. (prior art known as Quail+AMM~\cite{garimella2021sisyphus}) and this work (proposed retraining with second-order polynomial approximation of ReLU).}
\label{table:accuracy}
{\small
\begin{tabular*}{\columnwidth}
{@{}c@{}cccc} 
\toprule[1.5pt]
\multicolumn{1}{c}{\multirow{2}{*}[-0.65ex]{\parbox{1.4cm}{\centering \textbf{\textit{Dataset}}}}} 
& \multicolumn{1}{c}{\multirow{2}{*}[-0.65ex]{\parbox{1.8cm}{\centering \textbf{\textit{Network}}}}} 
& \multicolumn{1}{c}{\multirow{2}{*}[-0.65ex]{\parbox{1.2cm}{\centering \textbf{\textit{Base.}}}}}
& \multicolumn{1}{c}{\multirow{2}{*}[-0.65ex]{\parbox{1.3cm}{\centering \textbf{\textit{Prior.}}}}}
& \multicolumn{1}{c}{\textbf{\textit{This work}}}
\\
\cmidrule{5-5}
&&&& {$\bm{x^2+x}$}\\
\midrule[0.2pt]
\midrule[0.2pt]
C10 & VGG-16 & {94.66} & 82.25 & {91.80} \\ 
C10 & ResNet-18 & {95.87} & 83.61 & {91.44} \\ 
C10 & ResNet-32 & {92.66} & 56.93 & {87.56} \\ 
\midrule
C100 & VGG-16 & {74.91} & 54.56  & {65.57} \\ 
C100 & ResNet-18 & {79.49} & 65.17  & {73.25} \\ 
C100 & ResNet-32 & {70.01} & 19.86  & {60.15} \\ 
\midrule
Tiny & AlexNet & {51.82} & 36.24  & {45.53} \\ 
Tiny & VGG-11 & {57.81} & 44.63  & 50.17 \\ 
Tiny & VGG-16 & {61.02} & 45.76  & {54.83} \\ 
Tiny & ResNet-18 & {66.04} & 49.45  & {60.12} \\ 
Tiny & ResNet-32 & {49.18} & 7.16  & {39.97} \\ 
\bottomrule[1.5pt]
\end{tabular*}
}
\vspace{-1em}
\end{table}

In Flash, deep CNN models are trained using a nonlinear activation function $f(x) = x^2+x$, as proposed in \cite{ali2020polynomial}.
\ignore{\fixme{In addition to the advantages in accuracy over $x^2$}, t}The intuition behind using $x^2+x$ as the activation function is illustrated in Fig.~\ref{fig:relu}.
We note that $2\times\mathrm{ReLU}(x)$, or $2\times\max(0,x)$\footnote{The scaling factor 2 can be cancelled by making the weights and biases half in the linear layer.}, can be represented as $|x|+x$
and that $|x|$ is similar to $x^2$ in the range $(-1+\epsilon, 1+\epsilon)$ for small $\epsilon > 0$.
This implies that ReLU can be approximated as $x^2+x$ if the activation inputs are distributed around 0, more specifically in the range of $(-1+\epsilon, 1+\epsilon)$.

While \cite{ali2020polynomial} exploits $x^2+x$ to train simpler networks, Flash introduces a new training strategy for constructing deep CNN models.
\ignore{Based on this observation, the proposed methodology to train deep CNN models is structured as follows.}
Initially, we construct networks employing all-ReLU to achieve high inference accuracy.
In these networks, batch normalization (BN) layers are placed between convolution and ReLU to normalize the convolution output values before they enter the activation function. 
These models are trained, establishing our baseline (see Table~\ref{table:accuracy}).

Next, we start substituting ReLU with $x^2+x$ in the baseline networks. 
ReLUs are not replaced all at once as the squaring term in the polynomial activation can lead to exponential growth in the hidden layers' output values.
Even with BN, after passing only 2-3 polynomial layers, the output range exceeds [-1, 1] (the range where ReLU and $x^2+x$ are similar), leading to ineffective learning. 
To prevent this, we add a trainable scale factor multiplied to the output after BN, adjusting the output range. 
In addition, instead of retraining the entire network with polynomial activations at once, we progressively replace ReLU layer-by-layer and train the network.
In other words, from the first layer we start substituting layers with the CONV + ReLU structure with CONV + trainable scale factor + polynomial activation ($x^2+x$). 
Layers not yet replaced by polynomial activation functions are kept frozen with their baseline weights during retraining.
(A figure visualizing the proposed training method can be found in Appendix~\ref{subsection:training}.)
It enables the gradual transition of all-ReLU to all polynomial activations.

\ignore{Lastly, we proceed to retrain the reconstructed polynomial networks, where the internal activations are multiplied by the determined scale factor. 
This approach aims to recover from the performance degradation induced by the PI-friendly reform. 
Through this process, we successfully acquire PI-friendly CNNs with low-polynomial activations that exhibit performance akin to their ReLU-based counterparts.
Table~\ref{table:accuracy} shows the resulting model accuracy compared with the baseline models with ReLU and prior art with activation function as a quadratic function~\cite{garimella2021sisyphus}. }

\ignore{To validate the effectiveness of the proposed strategy, 
w}We train the same baseline networks using the $x^2$ activation function~\cite{gilad2016cryptonets,brutzkus2019low,garimella2021sisyphus}, following the same layer-by-layer training strategy with the scale factor as Flash,
and we compare the inference accuracy between CNNs (VGG-16, ResNet-18, and ResNet-32) with the \ignore{conventional} activation $x^2$ and ones with \ignore{the proposed} activation $x^2+x$. 
Fig.~\ref{fig:accuracy} plots the accuracy of each model as the ReLU layers are progressively replaced. 
For CNNs shown in Fig.~\ref{fig:accuracy}, (a) ResNet-32 on CIFAR-100, (b) ResNet-18 on TinyImageNet, and (c) VGG-16 on TinyImageNet, we observe that 1) \ignore{\fixme{the proposed polynomial}} $x^2+x$ shows consistently better accuracy compared to \ignore{the conventional}$x^2$ activation function and 2) their accuracy difference becomes larger as the models get deeper.
CNNs with $x^2+x$\ignore{the proposed polynomial} have 11.6\,\%, 28.1\,\%, and 30.6\,\% higher accuracy compared to $x^2$\ignore{the conventional activation} for VGG-16, ResNet-18, and ResNet-32, respectively.
Moreover, it shows\ignore{CNNs with the proposed polynomial show} accuracy comparable to the all-ReLU baseline models, which implies that $x^2+x$ is more amenable to training deeper models for PI. 
More details on training networks, which involves comparison between polynomial activations of $x^2+x$ and $x^2$, can be found in {Appendix~\ref{subsection:training}}. 


The proposed \ignore{polynomial and }training method has been validated over various CNNs and dataset. 
The trained models' accuracy is compared with the state-of-the-art training technique using the quadratic function for activation~\cite{garimella2021sisyphus} in Table~\ref{table:accuracy}.
Consistently over all the tested models and datasets, the proposed \ignore{polynomial and }training method shows average accuracy improvement of 15.9\,\% over \cite{garimella2021sisyphus}, while achieving much better accuracy for deeper CNNs like ResNet-32.

%% file: Contents/5_mpc.tex
\section{Secure Polynomial Activation Evaluation}
\label{sec:mpc}

\ignore{In Section~\ref{sec:training}, we show that CNNs with the polynomial activation $x^2 + x$ can be trained with high accuracy.} 
Although replacing all ReLUs in CNNs to second-order polynomial greatly reduces the PI communication cost,
existing methods to securely evaluate a polynomial as described in Section~\ref{sec:5.1} cannot achieve the target end-to-end CNN inference latency, less than 1 minute.

\subsection{Secure Polynomial Evaluation with Existing Techniques}
\label{sec:5.1}

One way to securely evaluate a polynomial activation function is using HE~\cite{gilad2016cryptonets,hesamifard2017cryptodl,brutzkus2019low,lee2022low,al2020towards, chou2018faster}.
Since ciphertext multiplication is the most computationally expensive operator in HE,
choosing a proper low-degree polynomial is critical to reduce the end-to-end inference latency.    
As described in Section~\ref{subsec:existing}, however, HE-based polynomial evaluation (even for $x^2$) suffers from huge computational overhead, making the end-to-end PI latency unacceptably large.

In order to avoid HE-induced computational overhead, many works have exploited 2PC-based protocols~\cite{juvekar2018gazelle,choi2022impala,mishra2020delphi,liu2017oblivious} to process the nonlinear activation layer.
In order to optimize the performance, existing 2PC-based techniques have tried
1) to minimize the number of ReLUs since the communication cost and latency of evaluating ReLU using GC is much higher than those of polynomial evaluation and
2) to minimize the online inference latency by offloading the data-independent operations onto the offline phase.
In Flash, since all the ReLUs are replaced with $x^2+x$, one of the existing 2PC-based protocols such as \cite{mishra2020delphi,liu2017oblivious} could have been employed.
However, employing existing techniques suffers from a drawback that the offline communication and computation cost is still huge.
Note that the offline phase should be processed for every inference (since no secrets can be reused),
and the total communication cost per inference is still too high, 
which remains the bottleneck to adoption of PI.
In view of these drawbacks, a novel 2PC-based protocol to securely evaluate the nonlinear activation function $x^2+x$, which has low communication cost and computational overhead, is proposed in the following.

\subsection{Proposed Secure Computation of Polynomial Activation Function} \label{subsec:proposed_protocol}

The proposed 2PC-based protocol solves the issues described in Section~\ref{sec:5.1} and reduces the total communication cost per inference significantly.
Specifically, it consists of two rounds of communication between two parties and does not require any offline communication, unlike existing 2PC-based protocols.
In addition, in the proposed protocol, packed HE ciphertexts containing additive secret shares are transmitted between two parties to minimize communication overhead.

\begin{figure}[t]
\centering
\includegraphics[width=0.8\columnwidth]{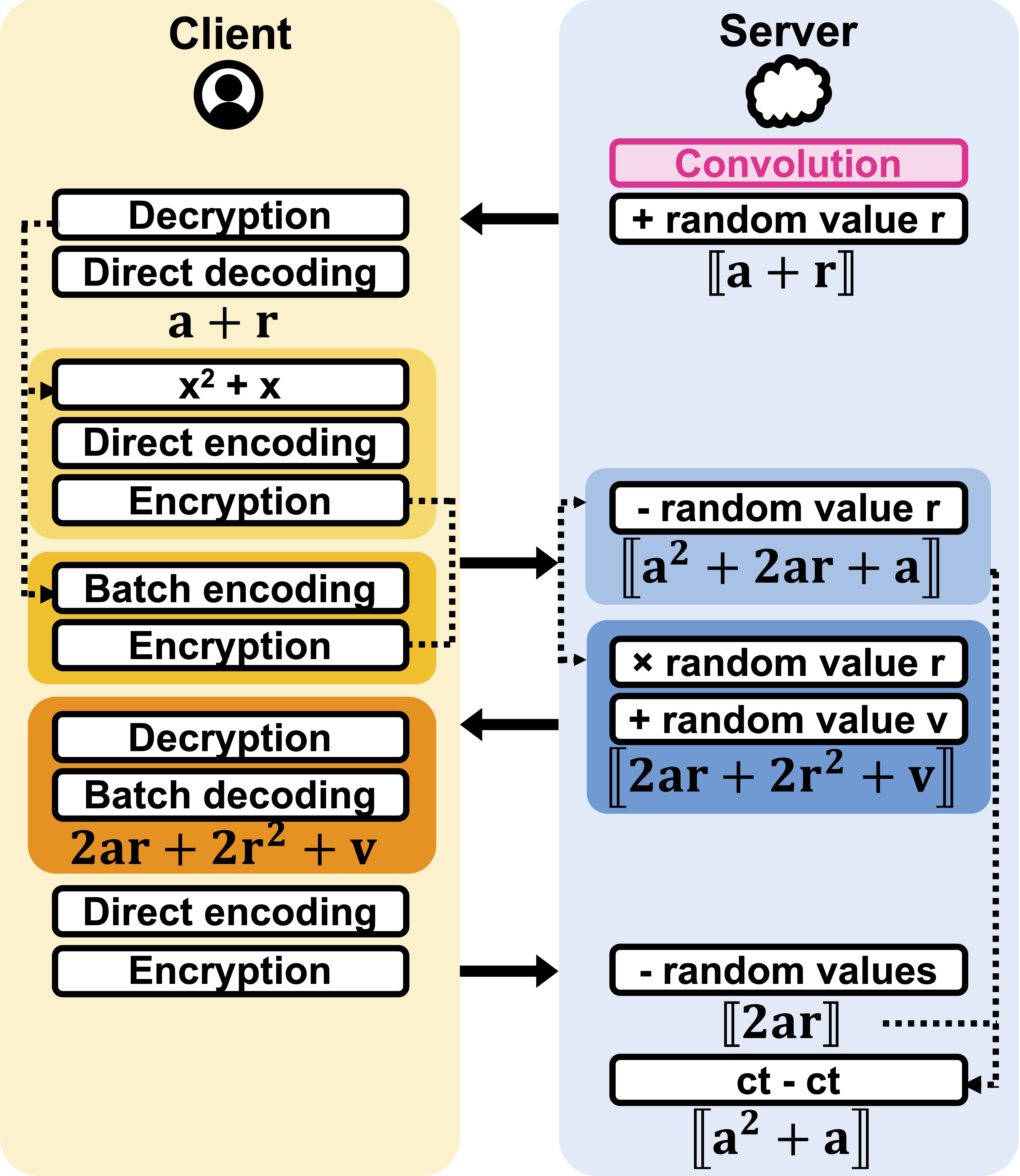}
\vspace{-0.5em}
\caption{Overall 2PC protocol for secure evaluation of $x^2+x$.}
\label{fig:proposed_mpc}
\vspace{-0.5em}
\end{figure}

The overall 2PC-based protocol that Flash uses to process the polynomial activation layer is described in Fig.~\ref{fig:proposed_mpc},
where the convolution output vector and the two random number vectors generated by the server are denoted as $\mathbf{a}$, $\mathbf{r}$, and $\mathbf{v}$, respectively.
After finishing convolution described in Section~\ref{sec:conv}, the server adds random numbers $\mathbf{r}$ to hide the convolution results in the ciphertext $\llbracket\mathbf{a}\rrbracket$, which can be performed by \textbf{HAdd($\llbracket\mathbf{a}\rrbracket$,$\llbracket\mathbf{r}\rrbracket$)}.
The result ciphertexts ($\llbracket\mathbf{a}+\mathbf{r}\rrbracket$ in Fig.~\ref{fig:proposed_mpc}) are then sent back to the client.
The client obtains the vectors of $\mathbf{a}+\mathbf{r}$ after decryption,
and the polynomial activation function $x^2+x$ is applied to each element of $\mathbf{a}+\mathbf{r}$. 
In other words, the client computes $\mathbf{a}^2+2\mathbf{a}\mathbf{r}+\mathbf{r^2}+\mathbf{a}+\mathbf{r}$, where arithmetic operations are applied element-wise.
Then, the client generates two different ciphertexts.
One ciphertext encrypts $\mathbf{a}^2+2\mathbf{a}\mathbf{r}+\mathbf{r^2}+\mathbf{a}+\mathbf{r}$ using the direct encoding described in Section~\ref{sec:conv},
and the other encrypts $\mathbf{a}+\mathbf{r}$ using the conventional batch encoding.
Both ciphertexts are transmitted to the server.

Now the server has $\mathbf{r}$, which is generated by the server itself, and two ciphertexts, $\llbracket\mathbf{a}+\mathbf{r}\rrbracket$ with batch encoding and $\llbracket\mathbf{a}^2+2\mathbf{a}\mathbf{r}+\mathbf{r^2}+\mathbf{a}+\mathbf{r}\rrbracket$ with direct encoding.
In order to obtain $\llbracket\mathbf{a}^2+\mathbf{a}\rrbracket$, which is the activation function output, the server needs to compute $\llbracket 2\mathbf{a}\mathbf{r}+\mathbf{r^2}+\mathbf{r}\rrbracket$.
Note that the server can obtain the ciphertext $\llbracket 2\mathbf{ar}+2\mathbf{r^2}\rrbracket$ using \textbf{PMult} since slot-wise SIMD multiplication between $2\mathbf{r}$ and $\llbracket\mathbf{a}+\mathbf{r}\rrbracket$ can be performed thanks to batch encoding.
However, we cannot apply homomorphic operations between $\llbracket 2\mathbf{ar}+2\mathbf{r^2}\rrbracket$ and $\llbracket\mathbf{a}^2+2\mathbf{a}\mathbf{r}+\mathbf{r^2}+\mathbf{a}+\mathbf{r}\rrbracket$ because they are encrypted with different encoding methods.

In order to solve this mismatch, the server adds new random numbers $\mathbf{v}$ to hide the intermediate results and sends the ciphertext $\llbracket 2\mathbf{ar}+2\mathbf{r^2}+\mathbf{v}\rrbracket$ to the client.
Then the client simply decrypts the ciphertext and encrypts $2\mathbf{ar}+2\mathbf{r^2}+\mathbf{v}$ using direct encoding, which is sent back to the server.
Now, since both ciphertexts $\llbracket\mathbf{a}^2+2\mathbf{a}\mathbf{r}+\mathbf{r^2}+\mathbf{a}+\mathbf{r}\rrbracket$ and $\llbracket2\mathbf{ar}+2\mathbf{r^2}+\mathbf{v}\rrbracket$ are encrypted using direct encoding,
the server can successfully compute $\llbracket\mathbf{a}^2+\mathbf{a}\rrbracket$ from both ciphertexts and the known random numbers $\mathbf{r}$ and $\mathbf{v}$.
Since the activation layer outputs are already direct-encoded in the output ciphertext,
convolution for the next layer can be performed subsequently.

\begin{table}
\begin{threeparttable}
\setlength{\tabcolsep}{4pt}
\centering
\caption{Amortized runtime and communication cost for evaluating individual GC-based ReLU, BT-based polynomial activation (Poly. act.) and proposed method in Flash.}
\label{table:relu}
{\small
\begin{tabular*}{\columnwidth}{@{\extracolsep{\fill}}@{}ccc@{}rcc}
\toprule[1.5pt]
\multicolumn{1}{c}{\multirow{2}{*}[-0.65ex]{\parbox{1.5cm}{\centering \textbf{\textit{Activation \\ function}}}}} & \multicolumn{2}{c}{\textbf{\textit{Offline}}} &
& \multicolumn{2}{c}{\textbf{\textit{Online}}} \\
\cmidrule{2-3}\cmidrule{5-6}
& {\textbf{Time ($\mathbf{\micro}\textrm{s}$)}} & {\textbf{Comm.}} & & {\textbf{Time ($\mathbf{\micro}\textrm{s}$)}} & {\textbf{Comm.}}\\
\midrule[0.2pt]
\midrule[0.2pt]
ReLU\footnotemark[1] & 60.60 & 19.1\,KB & & 20.22 & 1.184\,KB\\ [0.mm]
Poly. act.\footnotemark[1] & 2.80 & 0.192\,KB & & 1.20 & 0.036\,KB\\ [0.mm]
\textbf{Flash} & \textbf{0.21\footnotemark[2]} & \textbf{0 KB} & & \textbf{0.41} & \textbf{0.078\,KB} \\ [0.mm] 
\bottomrule[1.5pt]
\end{tabular*}
}
\begin{tablenotes}
\footnotesize
\item [1] Values are adopted from Table 1 in~\cite{park2022aespa}.
\item [2] This latency arises from generating random numbers $\mathbf{r}$ and $\mathbf{v}$.
\end{tablenotes}
\end{threeparttable}
\end{table}

Note that the proposed protocol does not require any offline communication and, thanks to packing and SIMD operation in HE, performing the protocol in Fig.~\ref{fig:proposed_mpc} once evaluates $n$ activation functions simultaneously, where $n$ is the HE parameter representing the polynomial degree (in Flash $n=2048$).
Hence, latency and communication cost to process the activation layer in Flash can be greatly improved, 
and its effectiveness compared to prior art is validated in Table~\ref{table:relu}.
While other 2PC-based techniques have large offline computation and communication cost to achieve good online performance, Flash does not need any offline communication. 
(The offline runtime in Table~\ref{table:relu} is to optimize the online latency, which will be described in more detail in Section~\ref{sec:offline}.)
Moreover, even the total latency and communication cost of the proposed protocol are lower than the online performance of prior art.

%% file: Contents/6_flash.tex
\section{Optimizing Flash for CPU Implementation}
\label{sec:flash}

In the following, several techniques to further optimize the implementation of Flash on CPU and to reduce the online end-to-end PI latency are introduced.

\subsection{Employing Lazy Reduction in Convolution}

To minimize the computational overhead due to modular multiplication during convolution, we employ lazy reduction.
Ciphertexts have 60-bit coefficients, and after being multiplied by kernel elements (\textbf{CMult}) ciphertexts as many as the number of kernel size and the number of input channels are added (see Algorithm~\ref{alg:conv}).
Even if 60-bit ciphertext coefficient magnitude grows by weight multiplication (less than 8-bit) and addition with hundreds of channels, 
the final result without reduction will still be much smaller than 120-bit. 
Although this lazy reduction requires two 64-bit words to represent each intermediate ciphertext coefficients and increases the latency for \textbf{HAdd} slightly,
bypassing reduction substantially decreases the total amount of computation,
which provides latency reduction by 2.2-2.5\,$\times$.
(Detailed data can be found in Appendix~\ref{subsec:conv}.)
Hence, rather than reducing after each multiplication and addition, we only reduce once after obtaining $\llbracket\mathbf{partial\ sum}\rrbracket$ in Algorithm~\ref{alg:conv}. 

\subsection{Accelerating Convolution with CPU Multi-threading}
The proposed convolution outperforms conventional one, especially when handling larger input sizes, as in Fig.~\ref{fig:conv_proposed}(c).
For instance, with an input size of 64$\times$64, we can obtain 56\,$\times$ speedup, 
but this performance gap decreases for the layers with smaller input sizes 
because this leads to more computation to produce output channels individually.
For a layer of $(H\times W, f_w, c_i, c_o) = (16^2, 3, 64, 64)$, it achieves only 12\,$\times$ speedup. 
While this is still a large improvement, the speedup is reduced by a factor of around 5. 
To accelerate convolution further, we harness the parallelism in convolution algorithm~\cite{reagen2021cheetah}. 
Specifically, the task for obtaining $\llbracket \mathbf{t}\rrbracket$ in Algorithm~\ref{alg:conv} can be parallelized and accelerated with CPU multi-threading.
Compared to the single-thread implementation using lazy reduction alone, multi-threading increases speed by 2.1-3.1\,$\times$ as detailed in Appendix~\ref{subsec:conv}.

\begin{algorithm}[t]
\caption{Online latency optimization for private-key BFV encryption.}
\label{alg:encryption}
\DontPrintSemicolon
\SetArgSty{textnormal}
\KwIn{
Plaintext $m(x)$ and ciphertext $\llbracket \mathbf{0}\rrbracket$
}
\KwOut{Ciphertext $\llbracket \mathbf{m}\rrbracket$}

    $\llbracket \mathbf{m}\rrbracket$=$\llbracket 
    \mathbf{0}\rrbracket$+$\Delta *m(x)$ \;
\end{algorithm}

\subsection{Moving Data-independent Operations to Offline Phase}
\label{sec:offline}

Many operations in the proposed protocol to evaluate $x^2+x$ are data-independent, which can be done during offline phase to reduce the online latency.
The random numbers $\mathbf{r}$ and $\mathbf{v}$ (see Fig.~\ref{fig:proposed_mpc}) generated by the server are data-independent, 
so Flash pre-generates and stores $\mathbf{r}$, $\mathbf{v}$, and even $\mathbf{r^2}$ in the offline phase to hide the latency due to random number generation. 
Runtime for generating random numbers is shown in Table~\ref{table:relu} under the offline latency of Flash.

Some of the client-side operations can be also done offline.
For private-key encryption in Table~\ref{table:enc_method}, only the term $\Delta m(x)$ is data-dependent and others are not,
so the client pre-computes $\llbracket\mathbf{0}\rrbracket$, which is ($[-({a(x)s(x)}+{e_0(x)})]_q$, $a(x)$), in the offline phase.
As in Algorithm~\ref{alg:encryption}, as soon as the client gets a message $\mathbf{m}$ for every layer, it encodes the message to plaintext, multiplies $m(x)$ with $\Delta$ and adds this product to the prepared $\llbracket\mathbf{0}\rrbracket$. 
With this approach, just scalar multiplication and addition are performed for encryption during online phase, resulting in an online latency of only 26.3\,$\micro\textrm{s}$ (9.4\,$\times$ speedup) as indicated in Table~\ref{table:enc_method}.



%% file: Contents/7_results.tex
\section{Experimental Results}
\label{sec:results}

\begin{figure*}[t]
\centering
\includegraphics[width=1.8\columnwidth]{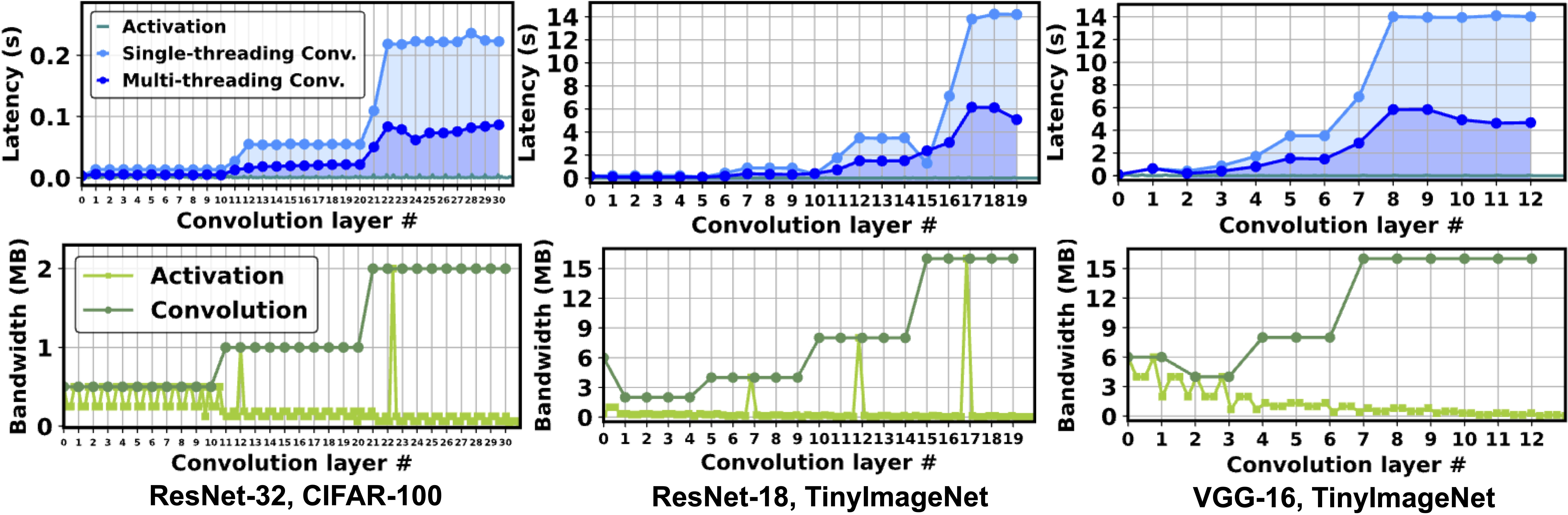}
\vspace{-0.5em}
\caption{Layer-by-layer breakdown of online latency and communication cost across various CNNs and datasets.}
\label{fig:layers}
\vspace{-0.5em}
\end{figure*}

Our experiments were conducted on workstations equipped with an Intel Xeon Gold 6250 CPU operating at 3.90\,GHz and 128\,GB of RAM, and the communication links between the parties are in the LAN setting similar to prior art. 
For PI benchmarks, we selected several standard CNN models, including ResNet-32~\cite{he2016deep} for CIFAR-100 (C100) and both ResNet-18~\cite{he2016deep} and VGG-16~\cite{simonyan2014very} for TinyImageNet (Tiny). 
The benchmark CNN architectures are described in Appendix~\ref{subsec:complexity}.
These models were implemented on CPU using C++ with SEAL~\cite{seal}. 
The encryption parameters were set to maintain a 128-bit security parameter~\cite{albrecht2015concrete,juvekar2018gazelle},\ignore{, consistent with the Gazelle~\cite{juvekar2018gazelle},} which employs a 19-bit plaintext and a 60-bit ciphertext modulus with a constant polynomial degree of 2048.

\subsection{Offline Communication Costs for Activation Layers}

Most prior art on PI focuses on improving online performance at the cost of increased offline overhead,
but if the communication cost in the offline phase is too high, it can significantly increase latency to the online phase due to bandwidth limitations~\cite{garimella2023characterizing}. 
Thus, optimizing the communication cost in both online and offline phases becomes indispensable.

Flash is the first hybrid PI protocol that eliminates the need for communication in the offline phase.
The hybrid protocols using GC such as \cite{juvekar2018gazelle,garimella2023characterizing} create and communicate GC during the offline phase, which incurs a substantial communication cost of 19.1\,KB per ReLU~\cite{park2022aespa}.
This leads to the latency equivalent to 11 minutes for models like ResNet-18~\cite{garimella2023characterizing}.
Even with methods optimized to reduce the cost of GC~\cite{choi2022impala, ghodsi2021circa} (Opt. GC), there is a communication cost of 3.7-4.5\,KB per ReLU. 
Compared to polynomial activation evaluation using BT proposed in \cite{mishra2020delphi,huang2022cheetah}, each square function evaluation costs 0.2\,KB~\cite{mishra2020delphi,park2022aespa}. 
These costs are summarized in Table~\ref{table:off_comm}, demonstrating that they require several GBs per inference for deep CNNs. 
In contrast, Flash completely eliminates any communication in the offline phase thanks to the technique proposed in Section~\ref{sec:mpc}.

\begin{table}
\setlength{\tabcolsep}{4pt}
\centering
\caption{Offline communication costs for all activation layers used in a single inference across multiple networks (R for ResNet and V for VGG): R-32 for CIFAR-100 and R-18, V-16 for TinyImageNet.}
\label{table:off_comm}
{\small
\begin{tabular*}{\columnwidth}{@{\extracolsep{\fill}}@{}c@{}cc@{}rc@{}rc@{}rc}
\toprule[1.5pt]
\multicolumn{2}{c}{\multirow{2}{*}{\diagbox[dir=NW]{\textbf{\textit{Models}}}{\textbf{\footnotesize per act.}}}} & \textbf{\textit{GC}} & & \textbf{\textit{Opt. GC}} & & \textbf{\textit{BT}} & & \textbf{\textit{Flash}} \\
\cmidrule{3-3}\cmidrule{5-5}\cmidrule{7-7}\cmidrule{9-9}
& & {\textbf{19.1\,KB}} & & {\textbf{3.7\,KB}} & & {\textbf{{0.2\,KB}}} & & {\textbf{0\,KB}}\\
\midrule[0.2pt]
\midrule[0.2pt]
\multicolumn{2}{c}{R-32 (C100)} & 5.5\,GB && 1\,GB && 0.06\,GB && \textbf{0} \\ 
\multicolumn{2}{c}{R-18 (Tiny)} & 41\,GB && 7.7\,GB && 0.41\,GB && \textbf{0} \\ 
\multicolumn{2}{c}{V-16 (Tiny)} & 20\,GB && 3.8\,GB && 0.2\,GB && \textbf{0} \\ 
\bottomrule[1.5pt]
\end{tabular*}
}
\vspace{-1em}
\end{table}

\subsection{End-to-end CNN Private Inference}
The end-to-end PI performance of Flash, i.e., online latency and online communication cost, is evaluated.
In Fig.~\ref{fig:layers}, we depict the online latency and communication costs associated with the linear and nonlinear operators that occur across multiple networks and datasets layer-by-layer.
It is noteworthy that the time spent processing convolutions significantly outweighs the time for handling nonlinear activation layers. 
Across all networks, the most time-consuming layers takes 14 seconds.
These are typically located in the latter parts of a network where the input size is small.

\ignore{However, even in the slowest layer of $(H\times W, f_w, c_i, c_o)=(4^2, 3, 512, 512)$, our method is 1.35$\times$ faster than conventional methods, which take approximately 19.1 seconds.
The speedup is particularly notable in the early layers of the network, where the large input size eliminates the necessity for output packing. 
For example, layers with an input size of 64 exhibit a speedup of 48.4-205.3$\times$ compared to conventional implementations. 
Overall, the processing time in the case of conventional methods can be between 1.35-205.3$\times$ slower than our proposed approach across all convolution layers, and thus these timescales have been omitted from Fig.~\ref{fig:layers} for clarity.}

As described in Section~\ref{sec:flash}, Flash takes advantage of CPU multi-threading to accelerate convolution and to combat the associated slowdown for the layers where the input size is less than $n$.
This enhances 2.06-3.05\,$\times$ speedup compared to single-threading as depicted in Fig.~\ref{fig:layers}.
The processing time for the conventional convolution is found to be 3.8-93.6\,$\times$ slower than the proposed approach across all convolution layers. 
These timescales have been omitted from Fig.~\ref{fig:layers} for clarity.
Detailed data and comparison can be found in Appendix~\ref{subsec:conv}.

\ignore{This final latency represents a \fixme{2-4-fold} improvement when compared with prior methodologies.}

In case of the online communication cost, thanks to the proposed 2PC protocol in Section~\ref{sec:mpc}, the overhead incurred by processing the nonlinear activation layers is not significant. 
As illustrated in Fig.~\ref{fig:layers} (see lower plots), sending the convolution output ciphertexts back to the client occupies the major portion of the total communication.
This is because the proposed convolution does not support output channel packing, while ciphertexts transmitted during evaluating activation functions contains multiple channels.

In ResNet, excluding the cases when the client transmits unpacked ciphertexts to the server for processing the residual paths (the server then receives this and adds it to later layer results), the communication costs at all layers remain within the convolution output ciphertext size. 
Consequently, the required bandwidth never exceeds the convolution outputs in each network, requiring only a maximum bandwidth of 16\,MB.
\ignore{Table~\ref{table:overall} illustrates the comprehensive communication costs for entire networks corresponding to online costs, indicating that all schemes fall under 0.22\,GB.}

\begin{table}
\begin{threeparttable}
\setlength{\tabcolsep}{4pt}
\centering
\caption{Performance comparison: time (in minutes) and communication costs (in\,GB) across multiple networks\ignore{ (R for ResNet, and V for VGG)}. Note that in Flash, latency for the linear layer is an online cost, while in the baseline, it has both online (on) and offline (off) phase.}
\label{table:overall}
{\small
\begin{tabular*}{\columnwidth}{C{0.9cm}C{1cm}r@{}C{0.77cm}C{0.77cm}r@{}cc@{}rC{0.77cm}C{0.67cm}}
\toprule[1.5pt]
\multicolumn{1}{c}{\multirow{2}{*}[-0.65ex]{\parbox{0.9cm}{\centering \textbf{\textit{Model}}}}} & {\multirow{2}{*}[-0.65ex]{\parbox{0.7cm}{\centering \textbf{\textit{Sys.}}}}} &
& \multicolumn{2}{c}{\textbf{\textit{Linear time}}} &
& \multicolumn{2}{c}{\textbf{\textit{Act. time}}} &
& \multicolumn{2}{c}{\textbf{\textit{Act. comm.}}} \\
\cmidrule{4-5}\cmidrule{7-8}\cmidrule(r{0.5em}){10-11}
&&&\textbf{Off.}&\textbf{On.} && \textbf{Off.} & \textbf{On.}
&& \textbf{Off.} & \textbf{On.} \\
\midrule[0.2pt]
\midrule[0.2pt]
\multirow{3}{*}{R-32\footnotemark[2]} & Delphi && 0.62 & 0.11 & & 0.31 & 0.1 & & 5.52 & 0.34\\
 & \textbf{Flash} && \textbf{0} & \textbf{0.02} & & \textbf{0.001} & \textbf{0.004} && \textbf{0} & \textbf{0.07}\\
  \cmidrule[0.05pt]{2-2}\cmidrule[0.05pt]{4-5}\cmidrule[0.05pt]{7-8}\cmidrule[0.05pt](r{0.5em}){10-11}
  & \textbf{\textit{gain}}&& \multicolumn{2}{c}{\textbf{45$\times$ (6.7$\times$)\footnotemark[1]}} & & \textbf{270$\times$} & \textbf{25.5$\times$} & & 
  \multicolumn{2}{c}{\textbf{84$\times$ (5.3$\times$)\footnotemark[1]}} \\ 
  \midrule
\multirow{3}{*}{R-18} & Delphi && 17.7 & 0.81 & & 2.25 & 0.75 & & 40.56 & 2.52\\
 & \textbf{Flash} && \textbf{0} & \textbf{0.48} & & \textbf{0.003} & \textbf{0.01} & & 
 \textbf{0} & \textbf{0.22}\\
  \cmidrule[0.05pt]{2-2}\cmidrule[0.05pt]{4-5}\cmidrule[0.05pt]{7-8}\cmidrule[0.05pt](r{0.5em}){10-11}
  & \textbf{\textit{gain}} && \multicolumn{2}{c}{\textbf{38$\times$ (1.7$\times$)\footnotemark[1]}}& & \textbf{\textbf{654$\times$}} & \textbf{81.6$\times$} && \multicolumn{2}{c}{\textbf{196$\times$ (12$\times$)\footnotemark[1]}} \\ 
  \midrule
\multirow{3}{*}{V-16} & Delphi && 8.32 & 0.4 & & 1.12 & 0.37 && 20.13 & 1.25 \\
 & \textbf{Flash} && \textbf{0} &\textbf{0.56} & & \textbf{0.008} & \textbf{0.02} && \textbf{0} & \textbf{0.21}\\
  \cmidrule[0.05pt]{2-2}\cmidrule[0.05pt]{4-5}\cmidrule[0.05pt]{7-8}\cmidrule[0.05pt](r{0.5em}){10-11}
  & \textbf{\textit{gain}} && \multicolumn{2}{c}{\textbf{16$\times$ (0.7$\times$)\footnotemark[1]}}& & \textbf{\textbf{140$\times$}} & \textbf{22.9$\times$} && \multicolumn{2}{c}{\textbf{102$\times$ (6$\times$)\footnotemark[1]}} \\ 
\bottomrule[1.5pt]
\end{tabular*}
}
\begin{tablenotes}
\footnotesize
\item [1] The gain is obtained by comparing the combined online and offline costs between Delphi and Flash. (The values inside the parentheses denote the online performance gain specifically.)
\item [2] ResNet-32 uses CIFAR-100, the rest use TinyImageNet.
\end{tablenotes}
\end{threeparttable}
\vspace{-1em}
\end{table}

\subsection{Performance Comparison}

For performance comparison, we consider a CNN PI protocol composed of conventional convolution using HE and nonlinear layers using GC, e.g., Gazelle~\cite{juvekar2018gazelle} or Delphi~\cite{mishra2020delphi}. 
When employing GC for ReLU operations, there are offline and online phases. 
As previously explained, the offline phase generates and communicates the GC, while the client evaluates the GC in the online phase. 
All the baseline latency and communication costs are obtained based on Table~\ref{table:relu}.

While Gazelle processes the linear layer online by computing convolution on the encrypted inputs, Delphi moves this intensive computing offline.
Even compared with Delphi's online performance, Flash provides generally better (or at least comparable in the worst case) online latency (see Table~\ref{table:overall}).
Here, we assume the all-ReLU network with the Delphi protocol as a baseline.

Table~\ref{table:overall} summarizes the cost-saving gains in both linear and non-linear operators. 
For Flash with ResNet-32 on CIFAR-100, the online latency is 0.0204 minutes, and the communication cost is 0.07\,GB.
The latency for computing convolutions is 0.02 minutes, approximately 38\,$\times$ less than the 0.62 minutes latency in the baseline linear layer.
Even if the majority portion of computation in this layer is moved offline, leaving only secret sharing online, there is still a 6.7\,$\times$ latency reduction.
Including the online latency of nonlinear layers, there is a 10.3\,$\times$ reduction in overall online latency.
The offline processing time for nonlinear layers, attributed to random number generation, is 270\,$\times$ faster than the conventional method for generating GC.
When comparing all offline and online latencies for an end-to-end network, it is {45.6}\,$\times$ faster than the baseline.

The online communication cost incurred by processing nonlinear layers also sees a 5.3\,$\times$ reduction. 
When accounting for the offline communication costs associated with transmitting GC, Flash achieves an 84\,$\times$ reduction in total communication cost.

Similar results are obtained for ResNet-18 and VGG-16 on TinyImageNet.
For ResNet-18, the online latency, including linear and non-linear layers, is 0.49 minutes, with an online communication cost of 0.22\,GB. 
For VGG-16, the online latency is 0.58 minutes, with an online communication cost of 0.21\,GB. 
It means a 38\,$\times$ improvement in total linear layer computations including the offline phase for ResNet-18 and a 16\,$\times$ improvement for VGG-16.
Even if they are handled offline, there is still a 3.2\,$\times$ and 1.3\,$\times$ speed enhancement in the total online latency for ResNet-18 and VGG-16, respectively.
Comparing all offline and online latencies for an end-to-end network, Flash is 43.6\,$\times$ and 17.4\,$\times$ faster, respectively.

The online communication costs are reduced by 12\,$\times$ and 6\,$\times$.
Taking into account the offline-phase costs, the total communication costs are dropped by 196\,$\times$ and 102\,$\times$, respectively. 

\begin{figure}[t]
\centering
\includegraphics[width=0.75\columnwidth]{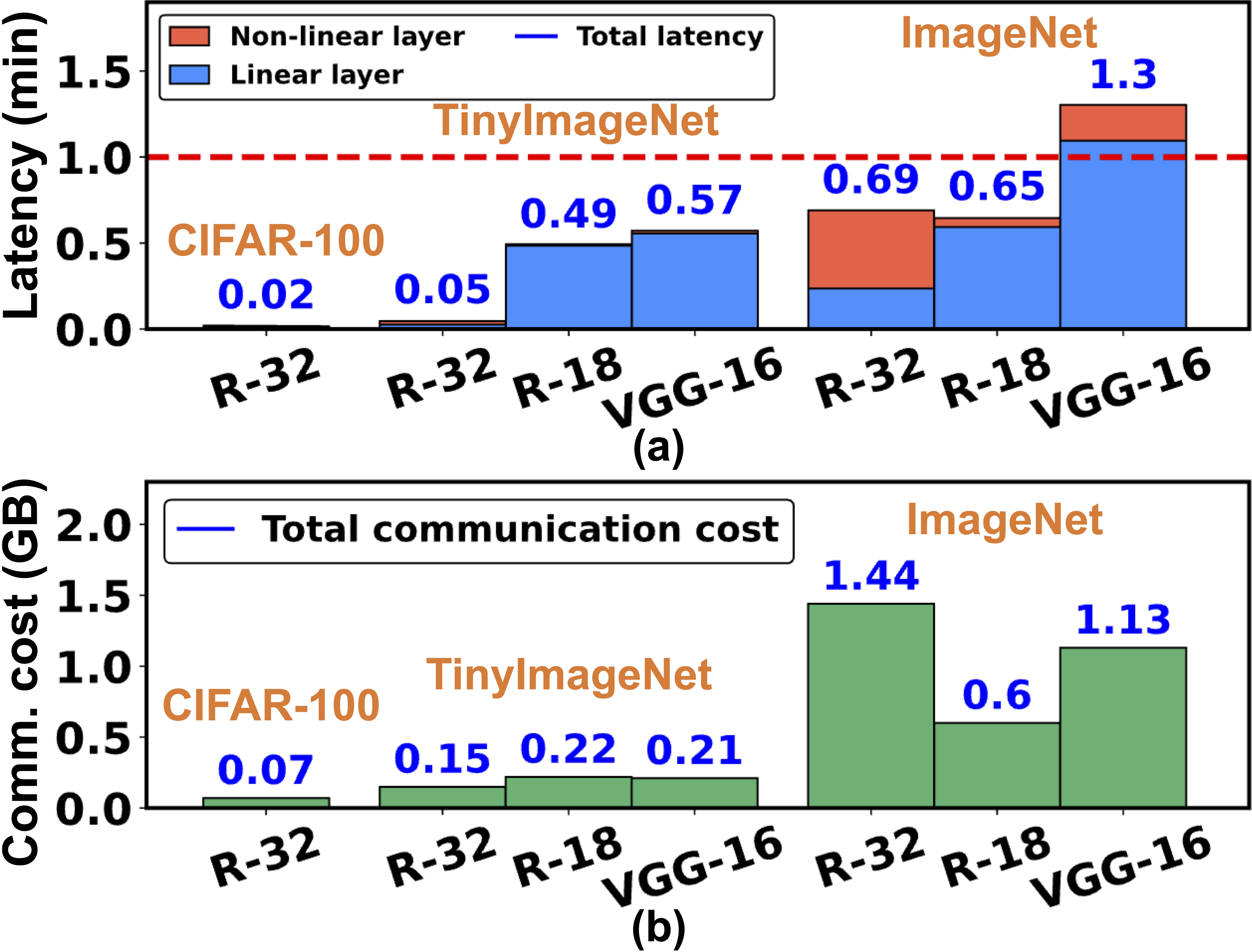}
\vspace{-1em}
\caption{Extended performance analysis for ResNet-32 on TinyImageNet and all networks on ImageNet: (a) end-to-end latency and (b) overall communication costs.}
\label{fig:imagenet}
\vspace{-0.9em}
\end{figure}

\subsection{Projected Impact on ImageNet Inference}

In PI, ImageNet is not commonly used for two main reasons. 
First, secure computation of ReLU is costly. 
Second, obtaining high inference accuracy is challenging. 
To elaborate the first issue, in the case of VGG-16, ResNet-32, and ResNet-18, processing ReLU through GC requires the storage and communication of 247\,GB, 271\,GB, and 498\,GB, respectively, in the offline phase. 
This volume of offline communication cost presents a significant burden given the available bandwidth, leading to processing ImageNet using BT in Squeezenet and ResNet-50 in \cite{huang2022cheetah,rathee2020cryptflow2}.

However, Flash requires no offline communication and demands less online bandwidth.
This can significantly mitigate the first issue. 
Therefore, we assume the completion of model training with polynomial activation for ImageNet, pre-construct Flash for the three networks, and measure the latency and communication cost.
We also incorporate the results for ResNet-32 on TinyImageNet.

As depicted in Fig.~\ref{fig:imagenet}, Flash can efficiently process various CNNs, achieving PI latency within a minute for all the networks except for VGG-16 on ImageNet. 
Latency ranges between 0.02 to 1.3 minutes along with the communication overhead of 0.07-1.44\,GB. 
These promising outcomes pave the way for future explorations into swift and lightweight PI for even more complex networks on large data.

%% file: Contents/9_conclusion.tex
\section{Conclusion}
\label{sec:conclusion}

This paper presents Flash, an optimized hybrid PI protocol utilizing both HE and 2PC, which can reduce the end-to-end PI latency for deep CNN models less than 1 minute on CPU.
To achieve this performance, first, Flash proposes a low-latency convolution algorithm implemented using a fast slot rotation \textbf{DRot} with a novel data vector encoding scheme, which results in {4-94}$\,\times$ performance improvement over the state-of-the-art.
Second, to minimize the communication cost introduced by ReLU,
Flash replaces the entire ReLUs with $x^2+x$ and trains deep CNN models with the new \ignore{activation function.} training strategy. 
The trained models improve the inference accuracy for CIFAR-10/100 and TinyImageNet by around 16\,\% on average (up to 40\,\% in ResNet-32) compared to prior art.
Last, Flash proposes a low-latency communication-efficient 2PC-based $x^2+x$ evaluation protocol that does not require any offline communication.
The proposed 2PC reduces the total communication cost to process the activation layer by {84-196}\,$\times$ over the state-of-the-art.
Flash optimized on CPU with these techniques achieves the end-to-end PI latency of 0.02 minute for CIFAR-100 and less than 0.57 minute for TinyImageNet, while the total data communication is 0.07\,GB for CIFAR-100 and less than 0.22\,GB for TinyImageNet.
Flash improves the state-of-the-art PI by {16-45}\,$\times$ in latency and {84-196}\,$\times$ in communication cost for the tested deep CNNs.
In addition, Flash can deliver the latency less than 1 minute on CPU with the total communication less than 1\,GB for ImageNet classification.

%% file: Contents/appendix.tex
\section*{Appendix}
\renewcommand{\thesubsection}{\Alph{subsection}}

\subsection{DRot with Direct Encoding}
\label{subsec:proof}

Flash takes advantage of the proposed \textbf{DRot} with direct encoding, which is by orders of magnitude faster than conventional \textbf{HRot} with batch encoding.
In the following, we prove that \textbf{DRot} with direct encoding performs slot rotation over encrypted data without key switching and that \textbf{DRot} does not add any noise in the ciphertext.

\noindent\textbf{Theorem.} Assume that a data vector is encoded to a plaintext using direct encoding as defined in \eqref{eq:de}
\begin{equation}
\mathbf{m} \mapsto m(x) = \sum_{i=0}^{n-1} \mathbf{m}[i]\cdot x^i, \nonumber
\end{equation}
and that $m(x)$ is encrypted using BFV to a ciphertext $\llbracket \mathbf{m}\rrbracket$=$(c_0(x), c_1(x))$.
Then, the following $\mathbf{DRot}(\llbracket \mathbf{m}\rrbracket,step)$ operation, as described in \eqref{eq:drot}, 
\begin{equation}
\mathrm{For}\, j \in \{0, 1\}, \quad c'_j (x) = c_j (x)\cdot x^{-step} \mod x^n+1, \nonumber
\end{equation}
returns a ciphertext $(c'_0(x), c'_1(x))$ that encrypts $\llbracket\langle \mathbf{m}\rangle_{step}\rrbracket$, i.e., \textbf{DRot} performs slot rotation over encrypted data.
In addition, noise in the output ciphertext $(c'_0(x), c'_1(x))$ after \textbf{DRot} is the same as that in the input $\llbracket \mathbf{m}\rrbracket$.

\noindent\textbf{Proof.} 
Since $\llbracket \mathbf{m}\rrbracket$=$(c_0(x), c_1(x))$ is encrypted with BFV, the ciphertext polynomials ($c_0(x)$ and $c_1(x)$) and the secret key $s(x)$ has the following relationship~\cite{fan2012somewhat}:
\begin{align}
\Delta m(x) + v(x)=[c_1(x)s(x)+c_0(x)]_q
\label{eq:bfv_dec}
\end{align}
for some noise polynomial $v(x)$. 
As long as $||v(x)||_\infty < \Delta/2$ (i.e., the largest magnitude of the coefficients of the noise polynomial $v(x)$ is smaller than $\Delta$/2), the message $m(x)$ can be correctly obtained after the BFV decryption process~\cite{fan2012somewhat}.

Multiplying $x^{-step}$ on both sides of \eqref{eq:bfv_dec} yields
\begin{align}
\Delta m(x)\cdot x^{-step} + v(x)\cdot x^{-step} = [c_1(x) \cdot x^{-step}s(x) + c_0(x) \cdot x^{-step}]_q \nonumber
\end{align}
and after taking (mod $x^n+1$) on both sides, we get
\begin{align}
[c'_1(x)s(x) + c'_0(x)]_q = \Delta m(x)\cdot x^{-step} + v(x)\cdot x^{-step} \mod x^n+1 \nonumber
\end{align}

The above equation implies that if the ciphertext $(c'_0(x), c'_1(x))$ is decrypted with the secret key $s(x)$, the plaintext $m(x)\cdot x^{-step}$ can be obtained.
Note that
\begin{align}
m(x)\cdot x^{-step} &=\sum_{i=0}^{n-1} \mathbf{m}[i]\cdot x^{i-step} \mod x^n+1 \nonumber \\
&= \sum_{i=0}^{n-1} (-1)^{\lfloor (i+step)/n \rfloor}\mathbf{m}[(i+step)\,\mathrm{mod}\,n]\cdot x^{i}, \nonumber
\end{align} 
which implies that the elements of the data vector $\mathbf{m}$ has undergone a left-cyclic shift by $step$.
Moreover, after $\mathbf{DRot}(\llbracket \mathbf{m}\rrbracket,step)$, the noise polynomial becomes $v(x)\cdot x^{-step} \mod x^n+1$, and it is obvious that $||v||_\infty = ||v(x)\cdot x^{-step}||_\infty$, so \textbf{DRot} does not add any noise in the ciphertext.  \hfill $\blacksquare$

\begin{figure}[t]
\centering
\includegraphics[width=1\columnwidth]{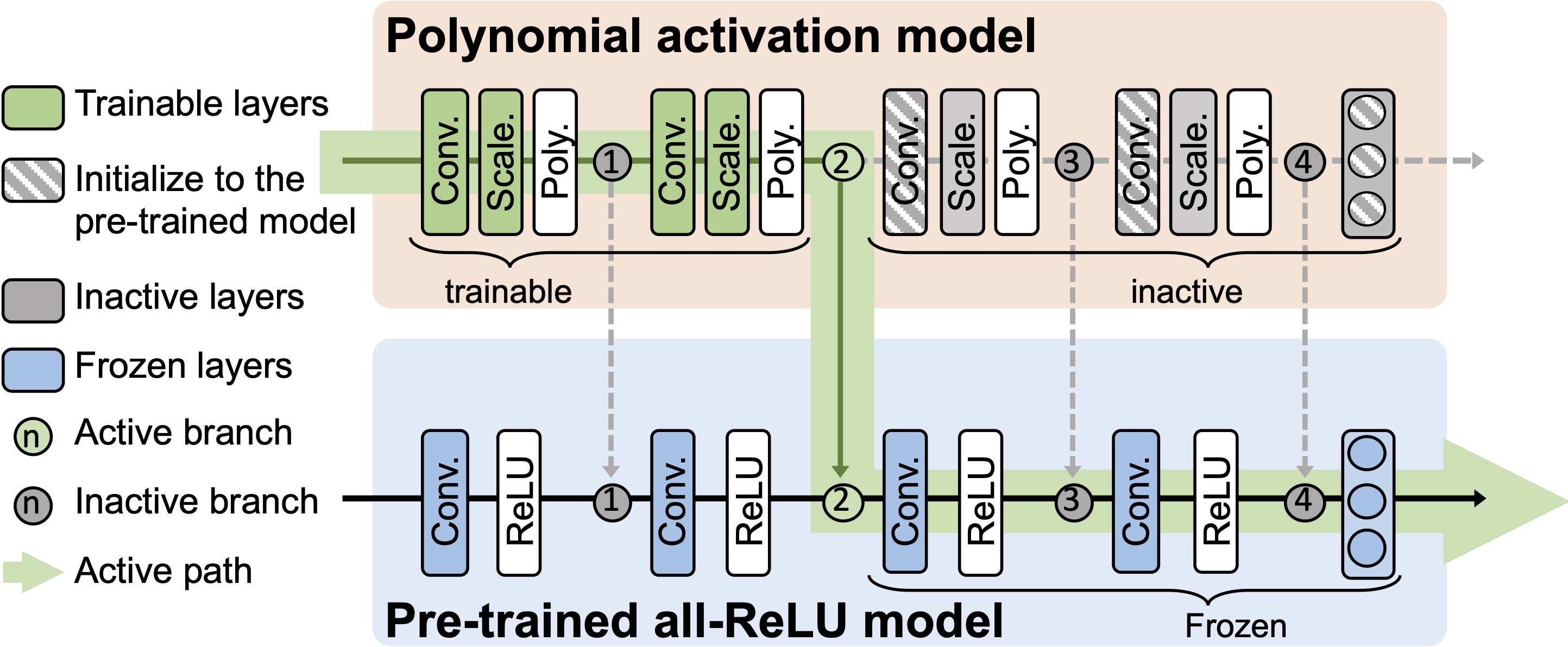}
\vspace{-0.5em}
\caption{Training strategy that gradually transitions from all-ReLU to all-poly activations.}
\label{fig:appendix_training}
\end{figure}

\begin{figure*}[t]
\centering
\includegraphics[width=1.85\columnwidth]{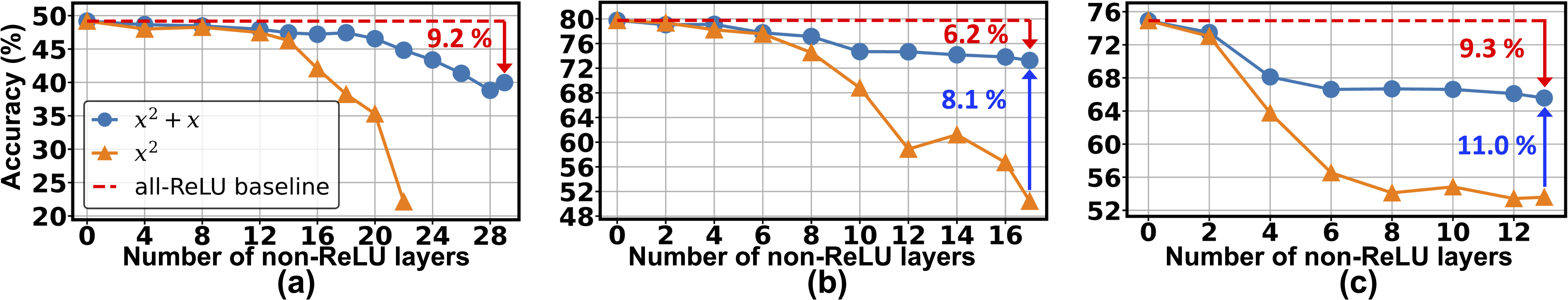}
\vspace{-0.5em}
\caption{Accuracy with replacement of ReLU: (a) ResNet-32 on TinyImageNet, (b) ResNet-18, and (c) VGG-16 on CIFAR-100.}
\label{fig:appendix_accuracy}
\vspace{-0.5em}
\end{figure*}

\subsection{Training Networks}
\label{subsection:training}
\noindent\textbf{Training strategy:}
Initially, a network with ReLU in all layers is set up to ensure high inference accuracy. 
Then, starting from the first layer, the existing CONV + ReLU structure is replaced with CONV + trainable scale factor + polynomial activation ($x^2+x$). 
During the retraining phase, layers still using ReLU remain frozen with their initial baseline weights.

For instance, consider a network with four convolution layers, as depicted in Fig.~\ref{fig:appendix_training}. 
The convolution weights of the polynomial activation model are initialized using the convolution weights from the pre-trained all-relu model. 
Then, each layer (or branch) is activated in sequence, and training proceeds layer-by-layer. 
Once training up to the fourth branch is complete, the entire network including the fully-connected layer is trained to finalize the all-polynomial activation model.

\noindent\textbf{Comparison between polynomial activations:}
Using the same setup and training strategy, we replicated the polynomial activation function $x^2$ as used in prior work~\cite{garimella2021sisyphus,mishra2020delphi,brutzkus2019low,ghodsi2020cryptonas,gilad2016cryptonets}, yielding the results shown in Table~\ref{table:appendix_accuracy}. 
Excluding ResNet-32, the average accuracy difference compared to state-of-the-art~\cite{garimella2021sisyphus} was within 7.7\,\%. 
The most significant difference was observed in ResNet-32, where a 29.3\,\% improvement in accuracy over previous research was seen on CIFAR-10, whereas on TinyImageNet, only 22 out of 29 ReLUs were substituted, not achieving full replacement. 
As Fig.~\ref{fig:accuracy} demonstrates accuracy changes across various datasets, the additional results are presented in Fig.~\ref{fig:appendix_accuracy}.
In Fig.~\ref{fig:appendix_accuracy}(a), for ResNet-32 on TinyImageNet, $x^2+x$ showed a 9.2\,\% lower result than the baseline. 
In (b), for ResNet-18 on CIFAR-100, $x^2+x$ was 6.2\,\% lower than the baseline and 8.1\,\% higher than $x^2$. 
In (c), for VGG-16 on CIFAR-100, the results were 9.3\,\% lower than the baseline and 11.0\,\% higher than $x^2$. 
Including this comparison, as demonstrated in Table~\ref{table:appendix_accuracy} showing an average 14\,\% higher accuracy, we determine that $x^2+x$ exhibits stronger convergence.

\begin{table}
\setlength{\tabcolsep}{4pt}
\centering
\caption{Test accuracy (\%) comparison on CIFAR-10/100 (C10/100) and TinyImageNet (Tiny) across multiple networks: Base. (all ReLU), all ReLU replaced with $x^2$, and this work ($x^2+x$), all evaluated under our experimental setup.}
\label{table:appendix_accuracy}
{\small
\begin{tabular*}{\columnwidth}
{@{}c@{}cccc} 
\toprule[1.5pt]
\multicolumn{1}{c}{\multirow{2}{*}[-0.65ex]{\parbox{1.4cm}{\centering \textbf{\textit{Dataset}}}}} 
& \multicolumn{1}{c}{\multirow{2}{*}[-0.65ex]{\parbox{1.8cm}{\centering \textbf{\textit{Network}}}}} 
& \multicolumn{1}{c}{\multirow{2}{*}[-0.65ex]{\parbox{1.1cm}{\centering \textbf{\textit{Base.}}}}}
& \multicolumn{1}{c}{\multirow{2}{*}[-0.65ex]{\parbox{1.2cm}{\centering \textbf{$x^2$}}}}
& \multicolumn{1}{c}{\textbf{\textit{This work}}}
\\
\cmidrule{5-5}
&&&& {$\bm{x^2+x}$}\\
\midrule[0.2pt]
\midrule[0.2pt]
C10 & VGG-16 & {94.66} & 88.66 & {91.80} \\ 
C10 & ResNet-18 & {95.87} & 70.49 & {91.44} \\ 
C10 & ResNet-32 & {92.66} & {86.24} & {87.56} \\ 
\midrule
C100 & VGG-16 & {74.91} & 53.59  & {65.57} \\ 
C100 & ResNet-18 & {79.49} & 50.43 & {73.25} \\ 
C100 & ResNet-32 & {70.01} & {29.56} & {60.15} \\ 
\midrule
Tiny & AlexNet & {51.82} & 41.27 & {45.53} \\ 
Tiny & VGG-11 & {57.81} & 43.72 & 50.17 \\ 
Tiny & VGG-16 & {61.02} & 43.19 & {54.83} \\ 
Tiny & ResNet-18 & {66.04} & 32.08 & {60.12} \\ 
Tiny & ResNet-32 & {49.18} & - & {39.97} \\ 
\bottomrule[1.5pt]
\end{tabular*}
}
\vspace{-1em}
\end{table}

\begin{table*}
\setlength{\tabcolsep}{4pt}
\centering
\caption{Comparison of runtime (in milliseconds) between conventional convolution and convolution with DRot, where $n$ is 2048. When $n$ is less than $H\times W$, multi-threading is applied.}
\label{table:appendix_conv}
{
\begin{tabular}{c@{}rccccccc}
\toprule[1.5pt]
\textit{\textbf{Parameter}} &
& \textit{\textbf{Filter width}}
& \multicolumn{1}{c}{\multirow{2}{*}[-0.4ex]{\parbox{2cm}{\centering \textbf{\textit{Conventional convolution}}}}}
& \multicolumn{3}{c}{\textbf{\textit{Convolution w/ DRot}}}
& \multicolumn{1}{c}{\multirow{2}{*}[-0.4ex]{\parbox{2cm}{\centering \textbf{\textit{Speedup}}}}}
\\
\cmidrule{1-1} \cmidrule{3-3} \cmidrule{5-7}
$(H\times W, c_i, c_o)$ && $f_w$& & \textbf{w/o lazy reduction} & \textbf{w/ lazy reduction} & \textbf{Multi-thread} & \\
\midrule[0.2pt]
\midrule[0.2pt]
($224^2,64,64$) && 3 & 425540.59 & 16756.19 & \textbf{7242.16} & - & $\bm{58.8\times}$ \\ [0.7mm]
($64^2,64,64$) && 3 & 35191.28 & 1499.09 & \textbf{631.79} & - & $\bm{55.7\times}$ \\ [0.7mm]
($64^2,3,64$) && 7 & 5111.23 & 377.3 & \textbf{173.95} & - & $\bm{29.4\times}$ \\ [0.7mm]
($56^2,256,256$) && 3 & 628739.03 & 16294.89 & \textbf{6667.16} & - & $\bm{94.3\times}$ \\ [0.7mm]
($32^2,16,16$) && 3 & 421.59 & 34.44 & 13.56 & \textbf{4.55} & $\bm{92.7\times}$ \\ [0.7mm]
($32^2,128,128$) && 3 & 31679.75 & 1975.86 & 866.76 & \textbf{403.66} & $\bm{78.5\times}$ \\ [0.7mm]
($28^2,512,512$) && 3 & 613657.28 & 32255.95 & 13526.23 & \textbf{6562.35} & $\bm{93.5\times}$ \\ [0.7mm]
($16^2,64,64$) && 3 & 2512.82 & 501.93 & 215.73 & \textbf{82} & $\bm{30.6\times}$ \\ [0.7mm]
($16^2,256,256$) && 3 & 48927.9 & 7921.42 & 3519.87 & \textbf{1469.13} & $\bm{33.3\times}$ \\ [0.7mm]
($8^2,128,128$) && 3 & 3112.02 & 1983.17 & 874.61 & \textbf{305.1} & $\bm{10.2\times}$ \\ [0.7mm]
($8^2,512,512$) && 3 & 73539.63 & 31684.55 & 14018.23 & \textbf{5831.25} & $\bm{12.6\times}$ \\ [0.7mm]
($4^2,512,512$) && 3 & 18477.85 & 31855.02 & 14103.22 & \textbf{4629.54} & $\bm{4\times}$ \\
\bottomrule[1.5pt]
\end{tabular}
}
\vspace{-1em}
\end{table*}

\subsection{Benchmark CNN Architectures}
\label{subsec:complexity}
In the following, we summarize the architectures of the networks used in Section~\ref{sec:results}.

\noindent\textbf{ResNet-32:}
It comprises of 33 convolution layers with a maximum filter size of $3\times3$ and maximum 64 output channels. After these layers, there are 31 ReLU activation layers because of two recurrent paths within the network. When using CIFAR-100, the total count of ReLU amounts to 303,104. Additionally, there is one average pooling layer with a size of eight in the context of CIFAR-100.

\noindent\textbf{ResNet-18:}
The network consists of 20 convolutional layers with a maximum filter size of $7\times7$ and up to 512 output channels. Following these layers, there are 17 ReLU activation layers due to the three recurrent paths. Using TinyImageNet as the dataset, the total count of ReLU activations reaches 2,228,224. There are two average pooling layers, one with a size of three and the other with a size of two.

\noindent\textbf{VGG-16:}
The network comprises 13 convolutional layers, each with a maximum filter size of $3\times3$ and a maximum of 512 output channels. 
The final three fully connected layers are treated as a single fully connected layer, and no ReLU activation function is applied to this layer.
Therefore, since ReLU is used only after each individual convolutional layer, there are a total of 13 ReLU layers in the entire network.
Using TinyImageNet, the total count of ReLU activations amounts to 1,105,920. 
There are five average pooling layers, each with a maximum size of four, within the context of TinyImageNet.

\begin{table}
\setlength{\tabcolsep}{4pt}
\centering
\caption{Comparison of the latency and communication costs between the proposed linear protocols and various protocols~\cite{rathee2020cryptflow2,huang2022cheetah}.}
\label{table:appendix_encoding}
{\small
\begin{tabular*}{\columnwidth}{cc@{}rccc}
\toprule[1.5pt]
\multicolumn{1}{c}{\multirow{2}{*}[-0.4ex]{\parbox{1.4cm}{\centering \textbf{\textit{Conv.}}}}}
& \textit{\textbf{Parameter}} &
& \textit{\textbf{Filter width}}
& \multicolumn{1}{c}{\multirow{2}{*}[-0.4ex]{\parbox{1.1cm}{\centering \textbf{\textit{LAN (s)}}}}}
& \multicolumn{1}{c}{\multirow{2}{*}[-0.4ex]{\parbox{1.1cm}{\centering \textbf{\textit{Commu. \\(MB)}}}}}
\\
\cmidrule{2-2} \cmidrule{4-4}
& $(H\times W, c_i, c_o)$ && $f_w$ && \\
\midrule[0.2pt]
\midrule[0.2pt]
\multirow{3}{*}[-0.4ex]{\parbox{1.4cm}{\centering CrypTFlow2\\ \cite{rathee2020cryptflow2}}} & ($224^2,3,64$) && 3 & 7.06 & 76.02 \\
& ($56^2,64,256$) && 1 & 8.21 & 28.01 \\
& ($56^2,256,64$) && 1 & 7.41 & 52.02 \\
\midrule[0.2pt]
\multirow{3}{*}[-0.4ex]{\parbox{1.4cm}{\centering Cheetah\\ \cite{huang2022cheetah}}} & ($224^2,3,64$) && 3 & 1.33 & 49.62 \\
& ($56^2,64,256$) && 1 & 0.83 & 15.3 \\
& ($56^2,256,64$) && 1 & 0.7 & 17.07 \\
\midrule[0.2pt]
\multirow{3}{*}{Flash} & ($224^2,3,64$) && 3 & 0.83	& 67 \\
& ($56^2,64,256$) && 1 & 0.33 & 20 \\
& ($56^2,256,64$) && 1 & 0.29 & 20 \\
\bottomrule[1.5pt]
\end{tabular*}
}
\vspace{-1em}
\end{table}

\subsection{Latency Comparison across Various Convolution Layers}
\label{subsec:conv}

\noindent\textbf{Comparison with the baseline:}
Table~\ref{table:appendix_conv} shows the runtimes for various convolution parameters of VGG-16 and ResNet-18/32, comparing conventional convolution and the proposed convolution with \textbf{DRot}.
Notably, lazy reduction has enhanced the speed of convolution with \textbf{DRot}, achieving a 2.2-2.5\,$\times$ reduction in latency.
For cases where the size of $H\times W$ is smaller than $n$ (here we set $n$ to 2048), 32 CPU multi-threading is applied to enhance the processing speed. 
Consequently, the results demonstrate a speed improvement ranging from 4-94\,$\times$ across different convolution parameters. 

\noindent\textbf{Comparison with other protocols:}
Table~\ref{table:appendix_encoding} provides comparison with convolution runtime implemented using HE in other protocols, specifically CrypTFlow2~\cite{rathee2020cryptflow2} and Cheetah~\cite{huang2022cheetah}. 
CrypTFlow2, like the baseline Delphi, uses batch encoding to leverage SIMD operations. 
For various convolutions, Flash is shown to be 8.5-25.6\,$\times$ faster and 1.1-2.6\,$\times$ more communication-efficient. 
On the other hand, Cheetah employs coefficient encoding to avoid expensive slot rotations. 
Although the encoding method in Cheetah is similar to the direct encoding in Flash, the different implementation of rotation and convolution in Flash results in a 1.6-2.5\,$\times$ speed improvement.

%% file: main.bbl

\begin{thebibliography}{75}


\ifx \showCODEN    \undefined \def \showCODEN     #1{\unskip}     \fi
\ifx \showDOI      \undefined \def \showDOI       #1{#1}\fi
\ifx \showISBNx    \undefined \def \showISBNx     #1{\unskip}     \fi
\ifx \showISBNxiii \undefined \def \showISBNxiii  #1{\unskip}     \fi
\ifx \showISSN     \undefined \def \showISSN      #1{\unskip}     \fi
\ifx \showLCCN     \undefined \def \showLCCN      #1{\unskip}     \fi
\ifx \shownote     \undefined \def \shownote      #1{#1}          \fi
\ifx \showarticletitle \undefined \def \showarticletitle #1{#1}   \fi
\ifx \showURL      \undefined \def \showURL       {\relax}        \fi
\providecommand\bibfield[2]{#2}
\providecommand\bibinfo[2]{#2}
\providecommand\natexlab[1]{#1}
\providecommand\showeprint[2][]{arXiv:#2}

\bibitem[Abadi et~al\mbox{.}(2016)]%
        {abadi2016deep}
\bibfield{author}{\bibinfo{person}{Martin Abadi}, \bibinfo{person}{Andy Chu}, \bibinfo{person}{Ian Goodfellow}, \bibinfo{person}{H~Brendan McMahan}, \bibinfo{person}{Ilya Mironov}, \bibinfo{person}{Kunal Talwar}, {and} \bibinfo{person}{Li Zhang}.} \bibinfo{year}{2016}\natexlab{}.
\newblock \showarticletitle{Deep learning with differential privacy}. In \bibinfo{booktitle}{\emph{Proceedings of the 2016 ACM SIGSAC conference on computer and communications security}}. \bibinfo{pages}{308--318}.
\newblock


\bibitem[Adi(1979)]%
        {adi1979share}
\bibfield{author}{\bibinfo{person}{Shamir Adi}.} \bibinfo{year}{1979}\natexlab{}.
\newblock \showarticletitle{How to share a secret}.
\newblock \bibinfo{journal}{\emph{Commun. ACM}}  \bibinfo{volume}{22} (\bibinfo{year}{1979}), \bibinfo{pages}{612--613}.
\newblock


\bibitem[Al~Badawi et~al\mbox{.}(2020)]%
        {al2020towards}
\bibfield{author}{\bibinfo{person}{Ahmad Al~Badawi}, \bibinfo{person}{Chao Jin}, \bibinfo{person}{Jie Lin}, \bibinfo{person}{Chan~Fook Mun}, \bibinfo{person}{Sim~Jun Jie}, \bibinfo{person}{Benjamin Hong~Meng Tan}, \bibinfo{person}{Xiao Nan}, \bibinfo{person}{Khin Mi~Mi Aung}, {and} \bibinfo{person}{Vijay~Ramaseshan Chandrasekhar}.} \bibinfo{year}{2020}\natexlab{}.
\newblock \showarticletitle{Towards the alexnet moment for homomorphic encryption: Hcnn, the first homomorphic cnn on encrypted data with gpus}.
\newblock \bibinfo{journal}{\emph{IEEE Transactions on Emerging Topics in Computing}} \bibinfo{volume}{9}, \bibinfo{number}{3} (\bibinfo{year}{2020}), \bibinfo{pages}{1330--1343}.
\newblock


\bibitem[Al~Badawi and Polyakov(2023)]%
        {al2023demystifying}
\bibfield{author}{\bibinfo{person}{Ahmad Al~Badawi} {and} \bibinfo{person}{Yuriy Polyakov}.} \bibinfo{year}{2023}\natexlab{}.
\newblock \showarticletitle{Demystifying bootstrapping in fully homomorphic encryption}.
\newblock \bibinfo{journal}{\emph{Cryptology ePrint Archive}} (\bibinfo{year}{2023}).
\newblock


\bibitem[Albrecht et~al\mbox{.}(2015)]%
        {albrecht2015concrete}
\bibfield{author}{\bibinfo{person}{Martin~R Albrecht}, \bibinfo{person}{Rachel Player}, {and} \bibinfo{person}{Sam Scott}.} \bibinfo{year}{2015}\natexlab{}.
\newblock \showarticletitle{On the concrete hardness of learning with errors}.
\newblock \bibinfo{journal}{\emph{Journal of Mathematical Cryptology}} \bibinfo{volume}{9}, \bibinfo{number}{3} (\bibinfo{year}{2015}), \bibinfo{pages}{169--203}.
\newblock


\bibitem[Ali et~al\mbox{.}(2020)]%
        {ali2020polynomial}
\bibfield{author}{\bibinfo{person}{Ramy~E Ali}, \bibinfo{person}{Jinhyun So}, {and} \bibinfo{person}{A~Salman Avestimehr}.} \bibinfo{year}{2020}\natexlab{}.
\newblock \showarticletitle{On polynomial approximations for privacy-preserving and verifiable relu networks}.
\newblock \bibinfo{journal}{\emph{arXiv preprint arXiv:2011.05530}} (\bibinfo{year}{2020}).
\newblock


\bibitem[Beaver(1995)]%
        {beaver1995precomputing}
\bibfield{author}{\bibinfo{person}{Donald Beaver}.} \bibinfo{year}{1995}\natexlab{}.
\newblock \showarticletitle{Precomputing oblivious transfer}. In \bibinfo{booktitle}{\emph{Annual International Cryptology Conference}}. Springer, \bibinfo{pages}{97--109}.
\newblock


\bibitem[Bellare et~al\mbox{.}(2013)]%
        {bellare2013efficient}
\bibfield{author}{\bibinfo{person}{Mihir Bellare}, \bibinfo{person}{Viet~Tung Hoang}, \bibinfo{person}{Sriram Keelveedhi}, {and} \bibinfo{person}{Phillip Rogaway}.} \bibinfo{year}{2013}\natexlab{}.
\newblock \showarticletitle{Efficient garbling from a fixed-key blockcipher}. In \bibinfo{booktitle}{\emph{2013 IEEE Symposium on Security and Privacy}}. IEEE, \bibinfo{pages}{478--492}.
\newblock


\bibitem[Blakley(1979)]%
        {blakley1979safeguarding}
\bibfield{author}{\bibinfo{person}{George~Robert Blakley}.} \bibinfo{year}{1979}\natexlab{}.
\newblock \showarticletitle{Safeguarding cryptographic keys}. In \bibinfo{booktitle}{\emph{Managing Requirements Knowledge, International Workshop on}}. IEEE Computer Society, \bibinfo{pages}{313--313}.
\newblock


\bibitem[Boemer et~al\mbox{.}(2019)]%
        {boemer2019ngraph}
\bibfield{author}{\bibinfo{person}{Fabian Boemer}, \bibinfo{person}{Anamaria Costache}, \bibinfo{person}{Rosario Cammarota}, {and} \bibinfo{person}{Casimir Wierzynski}.} \bibinfo{year}{2019}\natexlab{}.
\newblock \showarticletitle{nGraph-HE2: A high-throughput framework for neural network inference on encrypted data}. In \bibinfo{booktitle}{\emph{Proceedings of the 7th ACM Workshop on Encrypted Computing \& Applied Homomorphic Cryptography}}. \bibinfo{pages}{45--56}.
\newblock


\bibitem[Bossuat et~al\mbox{.}(2021)]%
        {bossuat2021efficient}
\bibfield{author}{\bibinfo{person}{Jean-Philippe Bossuat}, \bibinfo{person}{Christian Mouchet}, \bibinfo{person}{Juan Troncoso-Pastoriza}, {and} \bibinfo{person}{Jean-Pierre Hubaux}.} \bibinfo{year}{2021}\natexlab{}.
\newblock \showarticletitle{Efficient bootstrapping for approximate homomorphic encryption with non-sparse keys}. In \bibinfo{booktitle}{\emph{Annual International Conference on the Theory and Applications of Cryptographic Techniques}}. Springer, \bibinfo{pages}{587--617}.
\newblock


\bibitem[Brakerski et~al\mbox{.}(2014)]%
        {brakerski2014leveled}
\bibfield{author}{\bibinfo{person}{Zvika Brakerski}, \bibinfo{person}{Craig Gentry}, {and} \bibinfo{person}{Vinod Vaikuntanathan}.} \bibinfo{year}{2014}\natexlab{}.
\newblock \showarticletitle{(Leveled) fully homomorphic encryption without bootstrapping}.
\newblock \bibinfo{journal}{\emph{ACM Transactions on Computation Theory (TOCT)}} \bibinfo{volume}{6}, \bibinfo{number}{3} (\bibinfo{year}{2014}), \bibinfo{pages}{1--36}.
\newblock


\bibitem[Brutzkus et~al\mbox{.}(2019)]%
        {brutzkus2019low}
\bibfield{author}{\bibinfo{person}{Alon Brutzkus}, \bibinfo{person}{Ran Gilad-Bachrach}, {and} \bibinfo{person}{Oren Elisha}.} \bibinfo{year}{2019}\natexlab{}.
\newblock \showarticletitle{Low latency privacy preserving inference}. In \bibinfo{booktitle}{\emph{International Conference on Machine Learning}}. PMLR, \bibinfo{pages}{812--821}.
\newblock


\bibitem[Chen and Zhao(2012)]%
        {chen2012data}
\bibfield{author}{\bibinfo{person}{Deyan Chen} {and} \bibinfo{person}{Hong Zhao}.} \bibinfo{year}{2012}\natexlab{}.
\newblock \showarticletitle{Data security and privacy protection issues in cloud computing}. In \bibinfo{booktitle}{\emph{2012 international conference on computer science and electronics engineering}}, Vol.~\bibinfo{volume}{1}. IEEE, \bibinfo{pages}{647--651}.
\newblock


\bibitem[Cheon et~al\mbox{.}(2017)]%
        {cheon2017homomorphic}
\bibfield{author}{\bibinfo{person}{Jung~Hee Cheon}, \bibinfo{person}{Andrey Kim}, \bibinfo{person}{Miran Kim}, {and} \bibinfo{person}{Yongsoo Song}.} \bibinfo{year}{2017}\natexlab{}.
\newblock \showarticletitle{Homomorphic encryption for arithmetic of approximate numbers}. In \bibinfo{booktitle}{\emph{Advances in Cryptology--ASIACRYPT 2017: 23rd International Conference on the Theory and Applications of Cryptology and Information Security, Hong Kong, China, December 3-7, 2017, Proceedings, Part I 23}}. Springer, \bibinfo{pages}{409--437}.
\newblock


\bibitem[Cho et~al\mbox{.}(2022a)]%
        {cho2022sphynx}
\bibfield{author}{\bibinfo{person}{Minsu Cho}, \bibinfo{person}{Zahra Ghodsi}, \bibinfo{person}{Brandon Reagen}, \bibinfo{person}{Siddharth Garg}, {and} \bibinfo{person}{Chinmay Hegde}.} \bibinfo{year}{2022}\natexlab{a}.
\newblock \showarticletitle{Sphynx: A Deep Neural Network Design for Private Inference}.
\newblock \bibinfo{journal}{\emph{IEEE Security \& Privacy}} \bibinfo{volume}{20}, \bibinfo{number}{5} (\bibinfo{year}{2022}), \bibinfo{pages}{22--34}.
\newblock


\bibitem[Cho et~al\mbox{.}(2022b)]%
        {cho2022selective}
\bibfield{author}{\bibinfo{person}{Minsu Cho}, \bibinfo{person}{Ameya Joshi}, \bibinfo{person}{Brandon Reagen}, \bibinfo{person}{Siddharth Garg}, {and} \bibinfo{person}{Chinmay Hegde}.} \bibinfo{year}{2022}\natexlab{b}.
\newblock \showarticletitle{Selective network linearization for efficient private inference}. In \bibinfo{booktitle}{\emph{International Conference on Machine Learning}}. PMLR, \bibinfo{pages}{3947--3961}.
\newblock


\bibitem[Choi et~al\mbox{.}(2022)]%
        {choi2022impala}
\bibfield{author}{\bibinfo{person}{Woo-Seok Choi}, \bibinfo{person}{Brandon Reagen}, \bibinfo{person}{Gu-Yeon Wei}, {and} \bibinfo{person}{David Brooks}.} \bibinfo{year}{2022}\natexlab{}.
\newblock \showarticletitle{Impala: Low-Latency, Communication-Efficient Private Deep Learning Inference}.
\newblock \bibinfo{journal}{\emph{arXiv preprint arXiv:2205.06437}} (\bibinfo{year}{2022}).
\newblock


\bibitem[Choi et~al\mbox{.}(2018)]%
        {choi2018guaranteeing}
\bibfield{author}{\bibinfo{person}{Woo-Seok Choi}, \bibinfo{person}{Matthew Tomei}, \bibinfo{person}{Jose Rodrigo~Sanchez Vicarte}, \bibinfo{person}{Pavan~Kumar Hanumolu}, {and} \bibinfo{person}{Rakesh Kumar}.} \bibinfo{year}{2018}\natexlab{}.
\newblock \showarticletitle{Guaranteeing local differential privacy on ultra-low-power systems}. In \bibinfo{booktitle}{\emph{2018 ACM/IEEE 45th Annual International Symposium on Computer Architecture (ISCA)}}. IEEE, \bibinfo{pages}{561--574}.
\newblock


\bibitem[Chou et~al\mbox{.}(2018)]%
        {chou2018faster}
\bibfield{author}{\bibinfo{person}{Edward Chou}, \bibinfo{person}{Josh Beal}, \bibinfo{person}{Daniel Levy}, \bibinfo{person}{Serena Yeung}, \bibinfo{person}{Albert Haque}, {and} \bibinfo{person}{Li Fei-Fei}.} \bibinfo{year}{2018}\natexlab{}.
\newblock \showarticletitle{Faster cryptonets: Leveraging sparsity for real-world encrypted inference}.
\newblock \bibinfo{journal}{\emph{arXiv preprint arXiv:1811.09953}} (\bibinfo{year}{2018}).
\newblock


\bibitem[Cormode et~al\mbox{.}(2018)]%
        {cormode2018privacy}
\bibfield{author}{\bibinfo{person}{Graham Cormode}, \bibinfo{person}{Somesh Jha}, \bibinfo{person}{Tejas Kulkarni}, \bibinfo{person}{Ninghui Li}, \bibinfo{person}{Divesh Srivastava}, {and} \bibinfo{person}{Tianhao Wang}.} \bibinfo{year}{2018}\natexlab{}.
\newblock \showarticletitle{Privacy at scale: Local differential privacy in practice}. In \bibinfo{booktitle}{\emph{Proceedings of the 2018 International Conference on Management of Data}}. \bibinfo{pages}{1655--1658}.
\newblock


\bibitem[Costan and Devadas(2016)]%
        {costan2016intel}
\bibfield{author}{\bibinfo{person}{Victor Costan} {and} \bibinfo{person}{Srinivas Devadas}.} \bibinfo{year}{2016}\natexlab{}.
\newblock \showarticletitle{Intel SGX explained}.
\newblock \bibinfo{journal}{\emph{Cryptology ePrint Archive}} (\bibinfo{year}{2016}).
\newblock


\bibitem[Damg{\aa}rd et~al\mbox{.}(2012)]%
        {damgaard2012multiparty}
\bibfield{author}{\bibinfo{person}{Ivan Damg{\aa}rd}, \bibinfo{person}{Valerio Pastro}, \bibinfo{person}{Nigel Smart}, {and} \bibinfo{person}{Sarah Zakarias}.} \bibinfo{year}{2012}\natexlab{}.
\newblock \showarticletitle{Multiparty computation from somewhat homomorphic encryption}. In \bibinfo{booktitle}{\emph{Annual Cryptology Conference}}. Springer, \bibinfo{pages}{643--662}.
\newblock


\bibitem[Dathathri et~al\mbox{.}(2019)]%
        {dathathri2019chet}
\bibfield{author}{\bibinfo{person}{Roshan Dathathri}, \bibinfo{person}{Olli Saarikivi}, \bibinfo{person}{Hao Chen}, \bibinfo{person}{Kim Laine}, \bibinfo{person}{Kristin Lauter}, \bibinfo{person}{Saeed Maleki}, \bibinfo{person}{Madanlal Musuvathi}, {and} \bibinfo{person}{Todd Mytkowicz}.} \bibinfo{year}{2019}\natexlab{}.
\newblock \showarticletitle{CHET: an optimizing compiler for fully-homomorphic neural-network inferencing}. In \bibinfo{booktitle}{\emph{Proceedings of the 40th ACM SIGPLAN conference on programming language design and implementation}}. \bibinfo{pages}{142--156}.
\newblock


\bibitem[Demmler et~al\mbox{.}(2015)]%
        {demmler2015aby}
\bibfield{author}{\bibinfo{person}{Daniel Demmler}, \bibinfo{person}{Thomas Schneider}, {and} \bibinfo{person}{Michael Zohner}.} \bibinfo{year}{2015}\natexlab{}.
\newblock \showarticletitle{ABY-A framework for efficient mixed-protocol secure two-party computation.}. In \bibinfo{booktitle}{\emph{NDSS}}.
\newblock


\bibitem[Dwork et~al\mbox{.}(2014)]%
        {dwork2014algorithmic}
\bibfield{author}{\bibinfo{person}{Cynthia Dwork}, \bibinfo{person}{Aaron Roth}, {et~al\mbox{.}}} \bibinfo{year}{2014}\natexlab{}.
\newblock \showarticletitle{The algorithmic foundations of differential privacy}.
\newblock \bibinfo{journal}{\emph{Foundations and Trends{\textregistered} in Theoretical Computer Science}} \bibinfo{volume}{9}, \bibinfo{number}{3--4} (\bibinfo{year}{2014}), \bibinfo{pages}{211--407}.
\newblock


\bibitem[Fan and Vercauteren(2012)]%
        {fan2012somewhat}
\bibfield{author}{\bibinfo{person}{Junfeng Fan} {and} \bibinfo{person}{Frederik Vercauteren}.} \bibinfo{year}{2012}\natexlab{}.
\newblock \showarticletitle{Somewhat practical fully homomorphic encryption}.
\newblock \bibinfo{journal}{\emph{Cryptology ePrint Archive}} (\bibinfo{year}{2012}).
\newblock


\bibitem[Garimella et~al\mbox{.}(2023)]%
        {garimella2023characterizing}
\bibfield{author}{\bibinfo{person}{Karthik Garimella}, \bibinfo{person}{Zahra Ghodsi}, \bibinfo{person}{Nandan~Kumar Jha}, \bibinfo{person}{Siddharth Garg}, {and} \bibinfo{person}{Brandon Reagen}.} \bibinfo{year}{2023}\natexlab{}.
\newblock \showarticletitle{Characterizing and optimizing end-to-end systems for private inference}. In \bibinfo{booktitle}{\emph{Proceedings of the 28th ACM International Conference on Architectural Support for Programming Languages and Operating Systems, Volume 3}}. \bibinfo{pages}{89--104}.
\newblock


\bibitem[Garimella et~al\mbox{.}(2021)]%
        {garimella2021sisyphus}
\bibfield{author}{\bibinfo{person}{Karthik Garimella}, \bibinfo{person}{Nandan~Kumar Jha}, {and} \bibinfo{person}{Brandon Reagen}.} \bibinfo{year}{2021}\natexlab{}.
\newblock \showarticletitle{Sisyphus: A cautionary tale of using low-degree polynomial activations in privacy-preserving deep learning}.
\newblock \bibinfo{journal}{\emph{arXiv preprint arXiv:2107.12342}} (\bibinfo{year}{2021}).
\newblock


\bibitem[Gentry(2009)]%
        {gentry2009fully}
\bibfield{author}{\bibinfo{person}{Craig Gentry}.} \bibinfo{year}{2009}\natexlab{}.
\newblock \bibinfo{booktitle}{\emph{A fully homomorphic encryption scheme}}.
\newblock \bibinfo{publisher}{Stanford university}.
\newblock


\bibitem[Ghodsi et~al\mbox{.}(2021)]%
        {ghodsi2021circa}
\bibfield{author}{\bibinfo{person}{Zahra Ghodsi}, \bibinfo{person}{Nandan~Kumar Jha}, \bibinfo{person}{Brandon Reagen}, {and} \bibinfo{person}{Siddharth Garg}.} \bibinfo{year}{2021}\natexlab{}.
\newblock \showarticletitle{Circa: Stochastic relus for private deep learning}.
\newblock \bibinfo{journal}{\emph{Advances in Neural Information Processing Systems}}  \bibinfo{volume}{34} (\bibinfo{year}{2021}), \bibinfo{pages}{2241--2252}.
\newblock


\bibitem[Ghodsi et~al\mbox{.}(2020)]%
        {ghodsi2020cryptonas}
\bibfield{author}{\bibinfo{person}{Zahra Ghodsi}, \bibinfo{person}{Akshaj~Kumar Veldanda}, \bibinfo{person}{Brandon Reagen}, {and} \bibinfo{person}{Siddharth Garg}.} \bibinfo{year}{2020}\natexlab{}.
\newblock \showarticletitle{Cryptonas: Private inference on a relu budget}.
\newblock \bibinfo{journal}{\emph{Advances in Neural Information Processing Systems}}  \bibinfo{volume}{33} (\bibinfo{year}{2020}), \bibinfo{pages}{16961--16971}.
\newblock


\bibitem[Gilad-Bachrach et~al\mbox{.}(2016)]%
        {gilad2016cryptonets}
\bibfield{author}{\bibinfo{person}{Ran Gilad-Bachrach}, \bibinfo{person}{Nathan Dowlin}, \bibinfo{person}{Kim Laine}, \bibinfo{person}{Kristin Lauter}, \bibinfo{person}{Michael Naehrig}, {and} \bibinfo{person}{John Wernsing}.} \bibinfo{year}{2016}\natexlab{}.
\newblock \showarticletitle{Crypto{N}ets: Applying neural networks to encrypted data with high throughput and accuracy}. In \bibinfo{booktitle}{\emph{International conference on machine learning}}. PMLR, \bibinfo{pages}{201--210}.
\newblock


\bibitem[Goldreich(1998)]%
        {goldreich1998secure}
\bibfield{author}{\bibinfo{person}{Oded Goldreich}.} \bibinfo{year}{1998}\natexlab{}.
\newblock \showarticletitle{Secure multi-party computation}.
\newblock \bibinfo{journal}{\emph{Manuscript. Preliminary version}} \bibinfo{volume}{78}, \bibinfo{number}{110} (\bibinfo{year}{1998}), \bibinfo{pages}{1--108}.
\newblock


\bibitem[Halevi and Shoup(2014)]%
        {halevi2014algorithms}
\bibfield{author}{\bibinfo{person}{Shai Halevi} {and} \bibinfo{person}{Victor Shoup}.} \bibinfo{year}{2014}\natexlab{}.
\newblock \showarticletitle{Algorithms in helib}. In \bibinfo{booktitle}{\emph{Advances in Cryptology--CRYPTO 2014: 34th Annual Cryptology Conference, Santa Barbara, CA, USA, August 17-21, 2014, Proceedings, Part I 34}}. Springer, \bibinfo{pages}{554--571}.
\newblock


\bibitem[He et~al\mbox{.}(2016)]%
        {he2016deep}
\bibfield{author}{\bibinfo{person}{Kaiming He}, \bibinfo{person}{Xiangyu Zhang}, \bibinfo{person}{Shaoqing Ren}, {and} \bibinfo{person}{Jian Sun}.} \bibinfo{year}{2016}\natexlab{}.
\newblock \showarticletitle{Deep residual learning for image recognition}. In \bibinfo{booktitle}{\emph{Proceedings of the IEEE conference on computer vision and pattern recognition}}. \bibinfo{pages}{770--778}.
\newblock


\bibitem[Hesamifard et~al\mbox{.}(2017)]%
        {hesamifard2017cryptodl}
\bibfield{author}{\bibinfo{person}{Ehsan Hesamifard}, \bibinfo{person}{Hassan Takabi}, {and} \bibinfo{person}{Mehdi Ghasemi}.} \bibinfo{year}{2017}\natexlab{}.
\newblock \showarticletitle{Crypto{DL}: Deep neural networks over encrypted data}.
\newblock \bibinfo{journal}{\emph{arXiv preprint arXiv:1711.05189}} (\bibinfo{year}{2017}).
\newblock


\bibitem[Huang et~al\mbox{.}(2022)]%
        {huang2022cheetah}
\bibfield{author}{\bibinfo{person}{Zhicong Huang}, \bibinfo{person}{Wen-jie Lu}, \bibinfo{person}{Cheng Hong}, {and} \bibinfo{person}{Jiansheng Ding}.} \bibinfo{year}{2022}\natexlab{}.
\newblock \showarticletitle{Cheetah: Lean and fast secure two-party deep neural network inference}. In \bibinfo{booktitle}{\emph{31st USENIX Security Symposium (USENIX Security 22)}}. \bibinfo{pages}{809--826}.
\newblock


\bibitem[Ishai et~al\mbox{.}(2003)]%
        {ishai2003extending}
\bibfield{author}{\bibinfo{person}{Yuval Ishai}, \bibinfo{person}{Joe Kilian}, \bibinfo{person}{Kobbi Nissim}, {and} \bibinfo{person}{Erez Petrank}.} \bibinfo{year}{2003}\natexlab{}.
\newblock \showarticletitle{Extending oblivious transfers efficiently}. In \bibinfo{booktitle}{\emph{Annual International Cryptology Conference}}. Springer, \bibinfo{pages}{145--161}.
\newblock


\bibitem[Jha et~al\mbox{.}(2021)]%
        {jha2021deepreduce}
\bibfield{author}{\bibinfo{person}{Nandan~Kumar Jha}, \bibinfo{person}{Zahra Ghodsi}, \bibinfo{person}{Siddharth Garg}, {and} \bibinfo{person}{Brandon Reagen}.} \bibinfo{year}{2021}\natexlab{}.
\newblock \showarticletitle{Deepreduce: Relu reduction for fast private inference}. In \bibinfo{booktitle}{\emph{International Conference on Machine Learning}}. PMLR, \bibinfo{pages}{4839--4849}.
\newblock


\bibitem[Jung et~al\mbox{.}(2021)]%
        {jung2021over}
\bibfield{author}{\bibinfo{person}{Wonkyung Jung}, \bibinfo{person}{Sangpyo Kim}, \bibinfo{person}{Jung~Ho Ahn}, \bibinfo{person}{Jung~Hee Cheon}, {and} \bibinfo{person}{Younho Lee}.} \bibinfo{year}{2021}\natexlab{}.
\newblock \showarticletitle{Over 100x faster bootstrapping in fully homomorphic encryption through memory-centric optimization with gpus}.
\newblock \bibinfo{journal}{\emph{IACR Transactions on Cryptographic Hardware and Embedded Systems}} (\bibinfo{year}{2021}), \bibinfo{pages}{114--148}.
\newblock


\bibitem[Juvekar et~al\mbox{.}(2018)]%
        {juvekar2018gazelle}
\bibfield{author}{\bibinfo{person}{Chiraag Juvekar}, \bibinfo{person}{Vinod Vaikuntanathan}, {and} \bibinfo{person}{Anantha Chandrakasan}.} \bibinfo{year}{2018}\natexlab{}.
\newblock \showarticletitle{{GAZELLE}: A low latency framework for secure neural network inference}. In \bibinfo{booktitle}{\emph{27th USENIX Security Symposium (USENIX Security 18)}}. \bibinfo{pages}{1651--1669}.
\newblock


\bibitem[Kim et~al\mbox{.}(2023)]%
        {kim2023hyphen}
\bibfield{author}{\bibinfo{person}{Donghwan Kim}, \bibinfo{person}{Jaiyoung Park}, \bibinfo{person}{Jongmin Kim}, \bibinfo{person}{Sangpyo Kim}, {and} \bibinfo{person}{Jung~Ho Ahn}.} \bibinfo{year}{2023}\natexlab{}.
\newblock \showarticletitle{HyPHEN: A Hybrid Packing Method and Optimizations for Homomorphic Encryption-Based Neural Networks}.
\newblock \bibinfo{journal}{\emph{arXiv preprint arXiv:2302.02407}} (\bibinfo{year}{2023}).
\newblock


\bibitem[Kim et~al\mbox{.}(2022c)]%
        {kim2022ark}
\bibfield{author}{\bibinfo{person}{Jongmin Kim}, \bibinfo{person}{Gwangho Lee}, \bibinfo{person}{Sangpyo Kim}, \bibinfo{person}{Gina Sohn}, \bibinfo{person}{Minsoo Rhu}, \bibinfo{person}{John Kim}, {and} \bibinfo{person}{Jung~Ho Ahn}.} \bibinfo{year}{2022}\natexlab{c}.
\newblock \showarticletitle{Ark: Fully homomorphic encryption accelerator with runtime data generation and inter-operation key reuse}. In \bibinfo{booktitle}{\emph{2022 55th IEEE/ACM International Symposium on Microarchitecture (MICRO)}}. IEEE, \bibinfo{pages}{1237--1254}.
\newblock


\bibitem[Kim et~al\mbox{.}(2022a)]%
        {kim2022secure}
\bibfield{author}{\bibinfo{person}{Miran Kim}, \bibinfo{person}{Xiaoqian Jiang}, \bibinfo{person}{Kristin Lauter}, \bibinfo{person}{Elkhan Ismayilzada}, {and} \bibinfo{person}{Shayan Shams}.} \bibinfo{year}{2022}\natexlab{a}.
\newblock \showarticletitle{Secure human action recognition by encrypted neural network inference}.
\newblock \bibinfo{journal}{\emph{Nature communications}} \bibinfo{volume}{13}, \bibinfo{number}{1} (\bibinfo{year}{2022}), \bibinfo{pages}{4799}.
\newblock


\bibitem[Kim et~al\mbox{.}(2022b)]%
        {kim2022bts}
\bibfield{author}{\bibinfo{person}{Sangpyo Kim}, \bibinfo{person}{Jongmin Kim}, \bibinfo{person}{Michael~Jaemin Kim}, \bibinfo{person}{Wonkyung Jung}, \bibinfo{person}{John Kim}, \bibinfo{person}{Minsoo Rhu}, {and} \bibinfo{person}{Jung~Ho Ahn}.} \bibinfo{year}{2022}\natexlab{b}.
\newblock \showarticletitle{Bts: An accelerator for bootstrappable fully homomorphic encryption}. In \bibinfo{booktitle}{\emph{Proceedings of the 49th Annual International Symposium on Computer Architecture}}. \bibinfo{pages}{711--725}.
\newblock


\bibitem[Kundu et~al\mbox{.}(2023)]%
        {kundu2023learning}
\bibfield{author}{\bibinfo{person}{Souvik Kundu}, \bibinfo{person}{Shunlin Lu}, \bibinfo{person}{Yuke Zhang}, \bibinfo{person}{Jacqueline Liu}, {and} \bibinfo{person}{Peter~A Beerel}.} \bibinfo{year}{2023}\natexlab{}.
\newblock \showarticletitle{Learning to linearize deep neural networks for secure and efficient private inference}.
\newblock \bibinfo{journal}{\emph{arXiv preprint arXiv:2301.09254}} (\bibinfo{year}{2023}).
\newblock


\bibitem[Lee et~al\mbox{.}(2022)]%
        {lee2022low}
\bibfield{author}{\bibinfo{person}{Eunsang Lee}, \bibinfo{person}{Joon-Woo Lee}, \bibinfo{person}{Junghyun Lee}, \bibinfo{person}{Young-Sik Kim}, \bibinfo{person}{Yongjune Kim}, \bibinfo{person}{Jong-Seon No}, {and} \bibinfo{person}{Woosuk Choi}.} \bibinfo{year}{2022}\natexlab{}.
\newblock \showarticletitle{Low-complexity deep convolutional neural networks on fully homomorphic encryption using multiplexed parallel convolutions}. In \bibinfo{booktitle}{\emph{International Conference on Machine Learning}}. PMLR, \bibinfo{pages}{12403--12422}.
\newblock


\bibitem[Lee et~al\mbox{.}(2021)]%
        {lee2021high}
\bibfield{author}{\bibinfo{person}{Joon-Woo Lee}, \bibinfo{person}{Eunsang Lee}, \bibinfo{person}{Yongwoo Lee}, \bibinfo{person}{Young-Sik Kim}, {and} \bibinfo{person}{Jong-Seon No}.} \bibinfo{year}{2021}\natexlab{}.
\newblock \showarticletitle{High-precision bootstrapping of RNS-CKKS homomorphic encryption using optimal minimax polynomial approximation and inverse sine function}. In \bibinfo{booktitle}{\emph{Advances in Cryptology--EUROCRYPT 2021: 40th Annual International Conference on the Theory and Applications of Cryptographic Techniques, Zagreb, Croatia, October 17--21, 2021, Proceedings, Part I 40}}. Springer, \bibinfo{pages}{618--647}.
\newblock


\bibitem[Li et~al\mbox{.}(2020)]%
        {li2020federated}
\bibfield{author}{\bibinfo{person}{Tian Li}, \bibinfo{person}{Anit~Kumar Sahu}, \bibinfo{person}{Ameet Talwalkar}, {and} \bibinfo{person}{Virginia Smith}.} \bibinfo{year}{2020}\natexlab{}.
\newblock \showarticletitle{Federated learning: Challenges, methods, and future directions}.
\newblock \bibinfo{journal}{\emph{IEEE signal processing magazine}} \bibinfo{volume}{37}, \bibinfo{number}{3} (\bibinfo{year}{2020}), \bibinfo{pages}{50--60}.
\newblock


\bibitem[Liu et~al\mbox{.}(2017)]%
        {liu2017oblivious}
\bibfield{author}{\bibinfo{person}{Jian Liu}, \bibinfo{person}{Mika Juuti}, \bibinfo{person}{Yao Lu}, {and} \bibinfo{person}{Nadarajah Asokan}.} \bibinfo{year}{2017}\natexlab{}.
\newblock \showarticletitle{Oblivious neural network predictions via minionn transformations}. In \bibinfo{booktitle}{\emph{Proceedings of the 2017 ACM SIGSAC conference on computer and communications security}}. \bibinfo{pages}{619--631}.
\newblock


\bibitem[Lyubashevsky et~al\mbox{.}(2010)]%
        {lyubashevsky2010ideal}
\bibfield{author}{\bibinfo{person}{Vadim Lyubashevsky}, \bibinfo{person}{Chris Peikert}, {and} \bibinfo{person}{Oded Regev}.} \bibinfo{year}{2010}\natexlab{}.
\newblock \showarticletitle{On ideal lattices and learning with errors over rings}. In \bibinfo{booktitle}{\emph{Advances in Cryptology--EUROCRYPT 2010: 29th Annual International Conference on the Theory and Applications of Cryptographic Techniques, French Riviera, May 30--June 3, 2010. Proceedings 29}}. Springer, \bibinfo{pages}{1--23}.
\newblock


\bibitem[{Microsoft Research, Redmond, WA}(2023)]%
        {seal}
\bibfield{author}{\bibinfo{person}{{Microsoft Research, Redmond, WA}}.} \bibinfo{year}{2023}\natexlab{}.
\newblock \bibinfo{booktitle}{\emph{{SEAL. Simple Encrypted Arithmetic Library (release 4.1.1)}}}.
\newblock
\urldef\tempurl%
\url{https://github.com/microsoft/SEAL}
\showURL{%
\tempurl}


\bibitem[Mishra et~al\mbox{.}(2020)]%
        {mishra2020delphi}
\bibfield{author}{\bibinfo{person}{Pratyush Mishra}, \bibinfo{person}{Ryan Lehmkuhl}, \bibinfo{person}{Akshayaram Srinivasan}, \bibinfo{person}{Wenting Zheng}, {and} \bibinfo{person}{Raluca~Ada Popa}.} \bibinfo{year}{2020}\natexlab{}.
\newblock \showarticletitle{Delphi: A cryptographic inference system for neural networks}. In \bibinfo{booktitle}{\emph{Proceedings of the 2020 Workshop on Privacy-Preserving Machine Learning in Practice}}. \bibinfo{pages}{27--30}.
\newblock


\bibitem[Mohassel and Rindal(2018)]%
        {mohassel2018aby3}
\bibfield{author}{\bibinfo{person}{Payman Mohassel} {and} \bibinfo{person}{Peter Rindal}.} \bibinfo{year}{2018}\natexlab{}.
\newblock \showarticletitle{ABY3: A mixed protocol framework for machine learning}. In \bibinfo{booktitle}{\emph{Proceedings of the 2018 ACM SIGSAC conference on computer and communications security}}. \bibinfo{pages}{35--52}.
\newblock


\bibitem[Mohassel and Zhang(2017)]%
        {mohassel2017secureml}
\bibfield{author}{\bibinfo{person}{Payman Mohassel} {and} \bibinfo{person}{Yupeng Zhang}.} \bibinfo{year}{2017}\natexlab{}.
\newblock \showarticletitle{Secureml: A system for scalable privacy-preserving machine learning}. In \bibinfo{booktitle}{\emph{2017 IEEE symposium on security and privacy (SP)}}. IEEE, \bibinfo{pages}{19--38}.
\newblock


\bibitem[Mothukuri et~al\mbox{.}(2021)]%
        {mothukuri2021survey}
\bibfield{author}{\bibinfo{person}{Viraaji Mothukuri}, \bibinfo{person}{Reza~M Parizi}, \bibinfo{person}{Seyedamin Pouriyeh}, \bibinfo{person}{Yan Huang}, \bibinfo{person}{Ali Dehghantanha}, {and} \bibinfo{person}{Gautam Srivastava}.} \bibinfo{year}{2021}\natexlab{}.
\newblock \showarticletitle{A survey on security and privacy of federated learning}.
\newblock \bibinfo{journal}{\emph{Future Generation Computer Systems}}  \bibinfo{volume}{115} (\bibinfo{year}{2021}), \bibinfo{pages}{619--640}.
\newblock


\bibitem[Nilsson et~al\mbox{.}(2020)]%
        {nilsson2020survey}
\bibfield{author}{\bibinfo{person}{Alexander Nilsson}, \bibinfo{person}{Pegah~Nikbakht Bideh}, {and} \bibinfo{person}{Joakim Brorsson}.} \bibinfo{year}{2020}\natexlab{}.
\newblock \showarticletitle{A survey of published attacks on Intel SGX}.
\newblock \bibinfo{journal}{\emph{arXiv preprint arXiv:2006.13598}} (\bibinfo{year}{2020}).
\newblock


\bibitem[Park et~al\mbox{.}(2022)]%
        {park2022aespa}
\bibfield{author}{\bibinfo{person}{Jaiyoung Park}, \bibinfo{person}{Michael~Jaemin Kim}, \bibinfo{person}{Wonkyung Jung}, {and} \bibinfo{person}{Jung~Ho Ahn}.} \bibinfo{year}{2022}\natexlab{}.
\newblock \showarticletitle{AESPA: Accuracy preserving low-degree polynomial activation for fast private inference}.
\newblock \bibinfo{journal}{\emph{arXiv preprint arXiv:2201.06699}} (\bibinfo{year}{2022}).
\newblock


\bibitem[Patra and Suresh(2020)]%
        {patra2020blaze}
\bibfield{author}{\bibinfo{person}{Arpita Patra} {and} \bibinfo{person}{Ajith Suresh}.} \bibinfo{year}{2020}\natexlab{}.
\newblock \showarticletitle{BLAZE: blazing fast privacy-preserving machine learning}.
\newblock \bibinfo{journal}{\emph{arXiv preprint arXiv:2005.09042}} (\bibinfo{year}{2020}).
\newblock


\bibitem[Rathee et~al\mbox{.}(2020)]%
        {rathee2020cryptflow2}
\bibfield{author}{\bibinfo{person}{Deevashwer Rathee}, \bibinfo{person}{Mayank Rathee}, \bibinfo{person}{Nishant Kumar}, \bibinfo{person}{Nishanth Chandran}, \bibinfo{person}{Divya Gupta}, \bibinfo{person}{Aseem Rastogi}, {and} \bibinfo{person}{Rahul Sharma}.} \bibinfo{year}{2020}\natexlab{}.
\newblock \showarticletitle{Cryp{TF}low2: Practical 2-party secure inference}. In \bibinfo{booktitle}{\emph{Proceedings of the 2020 ACM SIGSAC Conference on Computer and Communications Security}}. \bibinfo{pages}{325--342}.
\newblock


\bibitem[Reagen et~al\mbox{.}(2021)]%
        {reagen2021cheetah}
\bibfield{author}{\bibinfo{person}{Brandon Reagen}, \bibinfo{person}{Woo-Seok Choi}, \bibinfo{person}{Yeongil Ko}, \bibinfo{person}{Vincent~T Lee}, \bibinfo{person}{Hsien-Hsin~S Lee}, \bibinfo{person}{Gu-Yeon Wei}, {and} \bibinfo{person}{David Brooks}.} \bibinfo{year}{2021}\natexlab{}.
\newblock \showarticletitle{Cheetah: Optimizing and accelerating homomorphic encryption for private inference}. In \bibinfo{booktitle}{\emph{2021 IEEE International Symposium on High-Performance Computer Architecture (HPCA)}}. IEEE, \bibinfo{pages}{26--39}.
\newblock


\bibitem[Riazi et~al\mbox{.}(2019)]%
        {riazi2019xonn}
\bibfield{author}{\bibinfo{person}{M~Sadegh Riazi}, \bibinfo{person}{Mohammad Samragh}, \bibinfo{person}{Hao Chen}, \bibinfo{person}{Kim Laine}, \bibinfo{person}{Kristin Lauter}, {and} \bibinfo{person}{Farinaz Koushanfar}.} \bibinfo{year}{2019}\natexlab{}.
\newblock \showarticletitle{$\{$XONN$\}$:$\{$XNOR-based$\}$ oblivious deep neural network inference}. In \bibinfo{booktitle}{\emph{28th USENIX Security Symposium (USENIX Security 19)}}. \bibinfo{pages}{1501--1518}.
\newblock


\bibitem[Riazi et~al\mbox{.}(2018)]%
        {riazi2018chameleon}
\bibfield{author}{\bibinfo{person}{M~Sadegh Riazi}, \bibinfo{person}{Christian Weinert}, \bibinfo{person}{Oleksandr Tkachenko}, \bibinfo{person}{Ebrahim~M Songhori}, \bibinfo{person}{Thomas Schneider}, {and} \bibinfo{person}{Farinaz Koushanfar}.} \bibinfo{year}{2018}\natexlab{}.
\newblock \showarticletitle{Chameleon: A hybrid secure computation framework for machine learning applications}. In \bibinfo{booktitle}{\emph{Proceedings of the 2018 on Asia conference on computer and communications security}}. \bibinfo{pages}{707--721}.
\newblock


\bibitem[Roh and Choi(2023)]%
        {roh2023hyena}
\bibfield{author}{\bibinfo{person}{Hyeri Roh} {and} \bibinfo{person}{Woo-Seok Choi}.} \bibinfo{year}{2023}\natexlab{}.
\newblock \showarticletitle{Hyena: Optimizing Homomorphically Encrypted Convolution for Private {CNN} Inference}.
\newblock \bibinfo{journal}{\emph{arXiv preprint arXiv:2311.12519}} (\bibinfo{year}{2023}).
\newblock


\bibitem[Samardzic et~al\mbox{.}(2021)]%
        {samardzic2021f1}
\bibfield{author}{\bibinfo{person}{Nikola Samardzic}, \bibinfo{person}{Axel Feldmann}, \bibinfo{person}{Aleksandar Krastev}, \bibinfo{person}{Srinivas Devadas}, \bibinfo{person}{Ronald Dreslinski}, \bibinfo{person}{Christopher Peikert}, {and} \bibinfo{person}{Daniel Sanchez}.} \bibinfo{year}{2021}\natexlab{}.
\newblock \showarticletitle{F1: A fast and programmable accelerator for fully homomorphic encryption}. In \bibinfo{booktitle}{\emph{MICRO-54: 54th Annual IEEE/ACM International Symposium on Microarchitecture}}. \bibinfo{pages}{238--252}.
\newblock


\bibitem[Satyanarayanan(2017)]%
        {satyanarayanan2017emergence}
\bibfield{author}{\bibinfo{person}{Mahadev Satyanarayanan}.} \bibinfo{year}{2017}\natexlab{}.
\newblock \showarticletitle{The emergence of edge computing}.
\newblock \bibinfo{journal}{\emph{Computer}} \bibinfo{volume}{50}, \bibinfo{number}{1} (\bibinfo{year}{2017}), \bibinfo{pages}{30--39}.
\newblock


\bibitem[Simonyan and Zisserman(2014)]%
        {simonyan2014very}
\bibfield{author}{\bibinfo{person}{Karen Simonyan} {and} \bibinfo{person}{Andrew Zisserman}.} \bibinfo{year}{2014}\natexlab{}.
\newblock \showarticletitle{Very deep convolutional networks for large-scale image recognition}.
\newblock \bibinfo{journal}{\emph{arXiv preprint arXiv:1409.1556}} (\bibinfo{year}{2014}).
\newblock


\bibitem[Smart and Vercauteren(2014)]%
        {smart2014fully}
\bibfield{author}{\bibinfo{person}{Nigel~P Smart} {and} \bibinfo{person}{Frederik Vercauteren}.} \bibinfo{year}{2014}\natexlab{}.
\newblock \showarticletitle{Fully homomorphic SIMD operations}.
\newblock \bibinfo{journal}{\emph{Designs, codes and cryptography}}  \bibinfo{volume}{71} (\bibinfo{year}{2014}), \bibinfo{pages}{57--81}.
\newblock


\bibitem[Subashini and Kavitha(2011)]%
        {subashini2011survey}
\bibfield{author}{\bibinfo{person}{Subashini Subashini} {and} \bibinfo{person}{Veeraruna Kavitha}.} \bibinfo{year}{2011}\natexlab{}.
\newblock \showarticletitle{A survey on security issues in service delivery models of cloud computing}.
\newblock \bibinfo{journal}{\emph{Journal of network and computer applications}} \bibinfo{volume}{34}, \bibinfo{number}{1} (\bibinfo{year}{2011}), \bibinfo{pages}{1--11}.
\newblock


\bibitem[Wagh et~al\mbox{.}(2019)]%
        {wagh2019securenn}
\bibfield{author}{\bibinfo{person}{Sameer Wagh}, \bibinfo{person}{Divya Gupta}, {and} \bibinfo{person}{Nishanth Chandran}.} \bibinfo{year}{2019}\natexlab{}.
\newblock \showarticletitle{SecureNN: 3-Party Secure Computation for Neural Network Training.}
\newblock \bibinfo{journal}{\emph{Proc. Priv. Enhancing Technol.}} \bibinfo{volume}{2019}, \bibinfo{number}{3} (\bibinfo{year}{2019}), \bibinfo{pages}{26--49}.
\newblock


\bibitem[Xu et~al\mbox{.}(2023)]%
        {xu2023falcon}
\bibfield{author}{\bibinfo{person}{Tianshi Xu}, \bibinfo{person}{Meng Li}, \bibinfo{person}{Runsheng Wang}, {and} \bibinfo{person}{Ru Huang}.} \bibinfo{year}{2023}\natexlab{}.
\newblock \showarticletitle{Falcon: Accelerating Homomorphically Encrypted Convolutions for Efficient Private Mobile Network Inference}. In \bibinfo{booktitle}{\emph{2023 IEEE/ACM International Conference on Computer Aided Design (ICCAD)}}. IEEE, \bibinfo{pages}{1--9}.
\newblock


\bibitem[Yao(1982)]%
        {yao1982protocols}
\bibfield{author}{\bibinfo{person}{Andrew~C Yao}.} \bibinfo{year}{1982}\natexlab{}.
\newblock \showarticletitle{Protocols for secure computations}. In \bibinfo{booktitle}{\emph{23rd annual symposium on foundations of computer science (sfcs 1982)}}. IEEE, \bibinfo{pages}{160--164}.
\newblock


\bibitem[Yao(1986)]%
        {yao1986generate}
\bibfield{author}{\bibinfo{person}{Andrew Chi-Chih Yao}.} \bibinfo{year}{1986}\natexlab{}.
\newblock \showarticletitle{How to generate and exchange secrets}. In \bibinfo{booktitle}{\emph{27th annual symposium on foundations of computer science (Sfcs 1986)}}. IEEE, \bibinfo{pages}{162--167}.
\newblock


\bibitem[Zahur et~al\mbox{.}(2015)]%
        {zahur2015two}
\bibfield{author}{\bibinfo{person}{Samee Zahur}, \bibinfo{person}{Mike Rosulek}, {and} \bibinfo{person}{David Evans}.} \bibinfo{year}{2015}\natexlab{}.
\newblock \showarticletitle{Two halves make a whole: Reducing data transfer in garbled circuits using half gates}. In \bibinfo{booktitle}{\emph{Advances in Cryptology-EUROCRYPT 2015: 34th Annual International Conference on the Theory and Applications of Cryptographic Techniques, Sofia, Bulgaria, April 26-30, 2015, Proceedings, Part II 34}}. Springer, \bibinfo{pages}{220--250}.
\newblock


\end{thebibliography}
